\documentclass[letterpaper]{article} 
\usepackage[draft]{aaai25}  
\usepackage{times}  
\usepackage{helvet}  
\usepackage{courier}  
\usepackage[hyphens]{url}  
\usepackage{graphicx} 
\usepackage{natbib}  
\usepackage{caption} 
\usepackage{algorithm}
\usepackage{algorithmic}

\usepackage{xcolor}
\newcommand{\answerYes}[1]{\textcolor{blue}{#1}} 
\newcommand{\answerNo}[1]{\textcolor{teal}{#1}} 
\newcommand{\answerNA}[1]{\textcolor{gray}{#1}} 

\newcommand{\revision}[1]{\textcolor{black}{#1}}
\newcommand{\revfig}[1]{#1} 

\usepackage{xurl}
\usepackage{hyperref}

\usepackage{subcaption}

\usepackage{newfloat}
\usepackage{listings}
\DeclareCaptionStyle{ruled}{labelfont=normalfont,labelsep=colon,strut=off} 
\floatstyle{ruled}
\newfloat{listing}{tb}{lst}{}
\floatname{listing}{Listing}
\title{Extracting Affect Aggregates from Longitudinal Social Media Data\\with Temporal Adapters for Large Language Models}
\author {
    Georg Ahnert\textsuperscript{\rm 1},
    Max Pellert\textsuperscript{\rm 2},
    David Garcia\textsuperscript{\rm 3,5},
    Markus Strohmaier\textsuperscript{\rm 1,4,5}
}
\affiliations {
    \textsuperscript{\rm 1}University of Mannheim,
    \textsuperscript{\rm 2}Barcelona Supercomputing Center,
    \textsuperscript{\rm 3}University of Konstanz,\\
    \textsuperscript{\rm 4}GESIS - Leibniz Institute for the Social Sciences
    \textsuperscript{\rm 5}CSH Vienna,
}

\begin{document}

\maketitle

\begin{figure*}[ht!] 
    \centering
    \includegraphics[width=\textwidth]{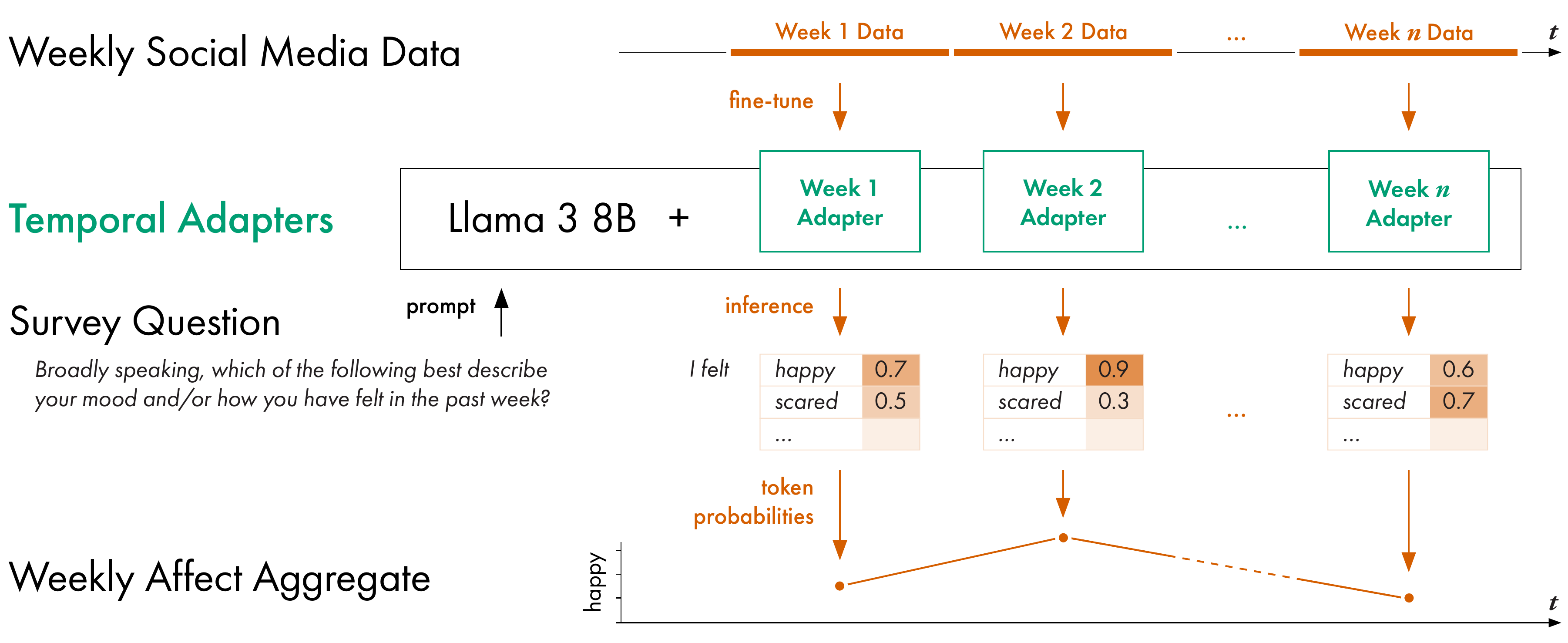} 
    \caption{\textbf{Illustration of Temporal Adapters.} First, we gather weekly text data from a panel of Twitter users and fine-tune Temporal Adapters for Llama 3 8B with it. Then, we prompt the fine-tuned model with established survey questions, one week at a time, and extract affect aggregates from the answer options' token probabilities. Temporal Adapters enable longitudinal analyses of affect aggregates from social media data by temporally aligning LLMs.}
    \label{fig:figure1}
\end{figure*}

\begin{abstract}
This paper proposes temporally aligned Large Language Models (LLMs) as a tool for longitudinal analysis of social media data.
We fine-tune Temporal Adapters for Llama 3 8B on full timelines from a panel of British Twitter users, and extract longitudinal aggregates of emotions and attitudes with established questionnaires.
\revision{We focus our analysis on the beginning of the COVID-19 pandemic that had a strong impact on public opinion and collective emotions.}
We validate our estimates against representative British survey data and find strong positive, significant correlations for several collective emotions.
The obtained estimates are robust across multiple training seeds and prompt formulations, and in line with collective emotions extracted using a traditional classification model trained on labeled data.
\revision{We demonstrate the flexibility of our method on questions of public opinion for which no pre-trained classifier is available.}
\revision{Our work extends the analysis of affect in LLMs to a longitudinal setting through Temporal Adapters. It enables flexible, new approaches towards the longitudinal analysis of social media data.}

\end{abstract}

%

\vspace{0.5cm}

\section{Introduction}

\paragraph{Problem}
Several recent studies have used Large Language Models (LLMs) to perform \textit{cross-sectional} surveys \textit{in silico}~\cite[e.g.,][]{argyle_out_2023, bisbee_synthetic_2023, durmus_towards_2023}, modeling human-based survey responses synthetically. 
While showing great promise, such prompt-based in silico surveys have not been aligned well from a temporal perspective, meaning that LLM training data and survey fieldwork periods (waves) typically did not meaningfully overlap temporally. Independently, many studies have investigated affect aggregates from social media data, in particular collective emotions~\cite[e.g.,][]{golder_diurnal_2011, pellert_validating_2022, metzler_collective_2023}. \revision{Research on collective emotions is important as it can provide actionable insights for public health messaging in times of crisis~\cite{lwin_global_2020} and on the social resilience of communities~\cite{garcia_collective_2019}.} However, these methods don't scale well as they typically rely on expensively curated dictionaries or on labeled training data.

\vspace{1cm}

\paragraph{Approach}
In this paper, we present a novel method for extracting longitudinal affect aggregates by using \textit{Temporal Adapters} fine-tuned on user-generated data from social media, as illustrated in Figure~\ref{fig:figure1}. \revision{Adapters allow researchers to efficiently fine-tune LLMs with data from new contexts by training only a very small fraction of additional parameters while keeping the original model weights frozen.} We base our analysis on a panel of 21,576 British Twitter users and obtain their full timelines from November 2019 to June 2020. First, we fine-tune Temporal Adapters for Llama 3 8B on 7-day subsets of the Twitter data. Second, we prompt the fine-tuned model with multiple established questionnaires~\cite{yougov_britains_2024, yougov_will_2024} and extract longitudinal affect aggregates from token probabilities. We evaluate our results against survey data collected from a representative sample of the British adult population and find strong positive and significant correlations for a subset of collective emotions. A comparison of results obtained using our method with a traditional classification model trained on labeled data shows strong agreement. We demonstrate robustness against prompt variation with an additional survey instrument that is constructed to measure the same affective phenomena~\cite{clark_panas-x_1994}. Experiments with synthetically mixed data indicate the internal validity of our method. Finally, we demonstrate that our flexible method can be used for the extraction of more complex collective attitudes as well.

\paragraph{Contribution}
The main contribution of this work is that it extends previous approaches to \textit{in silico} surveys with LLMs to a \textit{longitudinal} setting in which survey data and LLM training data are temporally aligned. Compared to previous methods that extract affect aggregates from user-generated data, our approach is not fixed to a specific survey question and neither relies on expensively curated dictionaries, nor on labeled data. \revision{Instead, our method allows researchers make use of the wealth of established survey questionnaires to extract affect aggregates from social media data.}
We provide a Python implementation under MIT license alongside our paper to facilitate replication and future work on extracting affect aggregates with LLMs.\footnote{\revision{\url{https://github.com/dess-mannheim/temporal-adapters}}}

\section{Related Work}

\subsection{Attitudes, Opinions, and Values in LLMs}

Many recent studies have investigated the application of cross-sectional surveys to LLMs~\cite[e.g.][]{argyle_out_2023, bisbee_synthetic_2023, atari_which_2023, von_der_heyde_vox_2023}.
Researchers typically prompt LLMs with survey questions from existing questionnaires that were originally developed to survey human populations~\cite{ma_potential_2024, agnew_illusion_2024}. This setup assumes that the diverse data on which LLMs were trained enables them to be viable proxies for population-level estimates of individual attributes. However, most research neglects the temporal alignment of training data and survey data, i.e., the fieldwork periods in which survey data was collected. We propose Temporal Adapters fine-tuned on user-generated data with a known date of creation to enable longitudinal alignment.

Research has shown that openly available, pretrained LLMs produce estimates that can be both culturally and politically biased when compared to population-level survey data~\cite[e.g.,][]{motoki_more_2023, santurkar_whose_2023, hartmann_political_2023, adilazuarda_towards_2024}.
Since our method is based on fine-tuning an existing LLM, we likely inherit some of these biases. Still, by fine-tuning the model on user-generated data, we have some degree of control the sampling process of whose attitudes, opinions, and values the model will learn.

LLMs were found to be sensitive to survey wording~\cite{tjuatja_llms_2024, dominguez-olmedo_questioning_2023, rottger_political_2024, mcilroy-young_set-based_2024}, and the particular answer scoring method~\cite{wang_my_2024, wang_look_2024}. We prompt each answer option separately to counteract order effects and we use an answer-prefix to be less dependent on first-token probabilities.
We demonstrate that our approach is robust with respect to specific prompt wording by extracting collective emotions with multiple survey instruments.

\paragraph{\revision{Temporal Adaptation of Language Models}}

\revision{Besides surveys, temporal adaptation of language models has been studied to combat language change over time. \citet{rottger_temporal_2021} used social media data to temporally adapt language models to both unlabeled and labeled training data. They found that, for document classification, temporal adaptation does not provide additional benefits to domain adaptation. Similarly, \citet{lazaridou_mind_2021} create a dynamic benchmark to account for changes in language when evaluating the performance of language models, and \citet{amba_hombaiah_dynamic_2021} propose incremental training regimes for language models. Instead of being concerned with general language model performance over time, our work focuses on using Temporal Adapters to extract affect aggregates from longitudinal social media data.}

\paragraph{Simulating Survey Participants with LLMs}

LLMs used for surveys are so far either studied as they were pre-trained, or they are fine-tuned on survey data~\cite{ramezani_knowledge_2023, kim_ai-augmented_2023}. With pre-trained models, researchers typically use \textit{silicon sampling} and prepend a diverse set of persona description, often derived from existing survey data, to the survey question~\cite{argyle_out_2023, sun_random_2024}. A major drawback of silicon sampling is that it is purely based on a model's pre-training data, potentially mis-portraying parts of society~\cite{wang_large_2024}. Our method is instead based around incorporating data generated by specific users at a specific time into LLMs that are then used for surveying.

\citet{kim_ai-augmented_2023} added embeddings obtained from LLMs to a neural network to predict longitudinal changes in opinions.
We instead work directly with a state-of-the-art LLM, to leverage its general language capabilities, and to extract affect aggregates using the same survey question that was asked to human survey participants. \citet{ramezani_knowledge_2023} trained an LLM on existing survey data and were able to improve similarity to human survey responses.
The training of Temporal Adapters is independent of existing survey data and allows us to extract affect aggregates from specific user-generated text datasets.

\subsection{Measuring Affect in Surveys}

\revision{With our approach, we extract affect aggregates from user-generated social media data. Our setup does not prescribe which exact survey question we use, flexibly allowing us to choose which construct specifically we want to study.}
In social psychology and sociology, \textit{affect} is used as the most general term that encompasses phenomena such as emotions and moods~\cite{mohiyeddini_what_2013, stets_measuring_2014}. \textit{Emotions} are typically more specific and short-lived, lasting between minutes and days, while \textit{moods} describe phenomena that are more diffuse and last between hours and weeks~\cite{oatley_cognitive_2014}, but the temporal distinction is not completely clear-cut.

\paragraph{Felt Emotions and Reactivity}
Researchers generally distinguish between \textit{felt} and \textit{expressed} emotions or moods~\cite{stets_measuring_2014}. Neither surveys nor social media data can tap into internal, felt emotions directly, so we focus only on expressed emotions in this paper. Survey measures of affect are further influenced by \textit{recall bias} and by reactivity because of their retrospective nature. \textit{Reactivity} refers to the fact that `the very knowledge that one is being observed can alter emotional experience'~\cite[p. 286]{stets_measuring_2014}.
Survey participants might also be more or less willing to report different emotions. Affect is an inherently social phenomena and central to interpersonal communication~\cite{lively_sociology_2016}, so social media platforms might intuitively seem like a suitable source of found data containing expressed emotions. In contrast to surveys, posting on social media can be more immediate and less distanced from the affective experience itself. Working with found data is also considerably cheaper than surveying. \revision{Disadvantages include that people on social media may behave vis-a-vis an audience, creating a source of performative bias.}

\subsection{Social Media Affect Macroscopes}

Text data generated by users on social media has been used to study election outcomes~\cite{gayo-avello_meta-analysis_2013} and consumer confidence~\cite{pasek_stability_2018}, with challenging results. One of the most common applications is the extraction of \textit{emotion macroscopes}~\cite{golder_diurnal_2011, garcia_social_2021, pellert_validating_2022}, in particular in response to catastrophic events~\cite{garcia_collective_2019, jones_this_2020} such as the beginning of the COVID-19 pandemic in 2020~\cite[e.g.,][]{valdez_social_2020, lwin_global_2020, ashokkumar_social_2021, metzler_collective_2023}. \revision{Collectively experienced emotions can be of different quality, magnitude, and duration than individual emotions~\cite{goldenberg_collective_2020} and research on collective emotions can provide actionable insights for public health messaging in times of crisis~\cite{lwin_global_2020} and on the social resilience of communities~\cite{garcia_collective_2019}.}
\revision{These methods are based on expensively created dictionaries or on models trained with labeled data~\cite[e.g.,][]{mohammad_semeval-2018_2018}, which are then typically used to detect positive/negative sentiment or individual emotions.}
Similar to our setup, estimates are aggregated for specific groups, or on a population level. \revision{In contrast to such earlier approaches, our method scales well by not relying on labeled training data or on expensively created dictionaries and can be flexibly applied to any survey question.}

\paragraph{Sampling Biases and Performative Behavior}

As mentioned before, social media data can suffer from sampling biases and consequences of performative behavior of users due to platform effects and community norms~\cite{sen_total_2021}. Estimates obtained from social media samples often do not generalize to a target population because of differences in internet penetration and platform use~\cite{hill_measuring_2020}. In addition, behavioral differences between groups lead to biases in representation~\cite{olteanu_social_2019}. In other words, different groups are more or less inclined to express emotions or attitudes on social media. Still, previous studies found strongly positive and robust correlations between collective emotions extracted from social media and survey responses on a population level~\cite{garcia_social_2021, pellert_validating_2022}. We extend this research by introducing an extraction method that is more flexible by extracting constructs directly through prompting LLMs with survey questions.

Neither self-reported affect in surveys, nor emotions extracted from user-generated social media data are by themselves perfectly accurate estimates. Since attitudes have a strong affective component~\cite{bergman_theoretical_1998}, attitudes suffer from similar biases in both survey data and social media data.
Our method, LLMs with Temporal Adapters, enables the flexible application of inference questions from the extensive, well-crafted collection of survey instruments for the extraction of longitudinal affect aggregates from social media data. Comparing population-level estimates obtained from both surveys and social media data will help us establish a more robust understanding of affective phenomena.

\section{Longitudinal Datasets}
We extract affect aggregates from Twitter panel data for Great Britain, covering 35 weeks from November 2019 to June 2020. This time frame includes both New Year 2020 and the first UK COVID-19 lockdown on March 23\textsuperscript{rd}, 2020. It allows us to investigate both seasonal patterns and a catastrophic event that had a large impact on emotions and attitudes, as identified in previous research. We decided against a larger time frame because of the large amount of computational resources required to fine-tune our model. \revision{The selected time frame follows previous research on extracting collective emotions from social media data during the COVID-19 pandemic~\cite[e.g.,][]{metzler_collective_2023}, but we would also like to emphasize the uniqueness of the situation. Substantive results such as cross-correlations between individual emotions might not generalize to other points in time, but drastically changing emotions offer a unique opportunity for our methodological research.}

\subsection{Twitter Panel Data} \label{subsec:twitter_panel}
We create a panel of in total 21,576 Twitter users (13.6 million tweets) with the following method: We use the commercial service Brandwatch to sample 10,000 tweets per day from accounts in Great Britain between the beginning of 2019 and March 2023. Brandwatch determines the location of users based on profile information and geo-tagged data~\cite{brandwatch_location_2020}. Next, we identify individual users in that sample and remove accounts that were classified as organizations, as well as accounts with too low or too high activity levels. We select the same number of users classified by Brandwatch as `female' and `male', based on self-reported profile information~\cite{brandwatch_forsight_2020}. We retrieve their full timelines of tweets including retweets and replies back to the beginning of 2019. On average, for the 35 weeks from November 2019 to June 2020, we have 385,000 tweets per week available.

\subsection{Questionnaires and Survey Data}
We extract affect aggregates by querying an LLM with existing survey questionnaires and compare our results to publicly available survey data that was collected using the same questions. The survey data was collected from a British online panel with a target population of all British adults aged 18+~\cite{yougov_panel_2024}. The panel provider uses active sampling and post-survey adjustment weights based on age, gender, social class, region, and level of education.

\paragraph{Britain's Mood, Measured Weekly} To extract collective emotions, we use the question developed by \citet{yougov_britains_2024}:
\begin{quote}
    Broadly speaking, which of the following best describe your mood and/or how you have felt in the past week?
\end{quote}
This question uses a multicode answer scale, that is, multiple of the following answer options can be selected: `happy', `sad', `energetic', `apathetic', `inspired', `frustrated', `optimistic', `stressed', `content', `bored', `lonely', and `scared'.
We assess our results by comparing them with aggregate survey data gathered from the aforementioned British panel on the same question. Survey data is available in weekly waves as aggregates over 1890 to 2081 participants per wave.


\paragraph{PANAS-X} We further investigate collective emotions using the extended version of the Positive and Negative Affect Schedule (PANAS-X) developed by \citet{clark_panas-x_1994}. This survey instrument asks participants to indicate to what extend they feel they have felt like a series of carefully compiled adjectives. It comes with a series of time instructions, ranging from how participants feel in the moment to in general. The instrument has been validated for \textit{state affect}, i.e., short-term fluctuations in mood, as well~\cite{clark_panas-x_1994}. We create a prompt based on the `week' instruction, to measure emotions in the same time frame as in \citeauthor{yougov_britains_2024}'s wording, as follows:
\begin{quote}
    To what extent have you felt [adjective] during the past week?
\end{quote}
We focus on two emotions included in both the PANAS-X and in \citeauthor{yougov_britains_2024}'s survey data, \textit{scared} and \textit{sad}. The answer options are `very slightly or not at all', `a little', `moderately', `quite a bit', and `extremely', to be answered for each adjective -- in contrast to \citeauthor{yougov_britains_2024}'s answer options, which are directly multiple choice among emotions. We extract answer probabilities for each adjective and each answer option, and combine all of them into a single score for each emotion, according to the instructions provided by \citet{clark_panas-x_1994}. We compare our results to survey data gathered by \citeauthor{yougov_britains_2024} in \textit{Britain's Mood, Measured Weekly} for the respective emotions.

\paragraph{\revision{Collective Public Opinion Beyond Emotions}}
\revision{In addition to collective emotions, we include 3 public opinion questions into our analysis to demonstrate the flexibility of our approach. We select these questions based on available YouGov data for evaluation and based on their relevance to the beginning of the COVID-19 pandemic, during which the UK government had to implement stringent measures with large consequences on people's everyday life.}
We extract collective attitudes towards the National Health Service (NHS) with the following question~\cite{yougov_will_2024}:
\begin{quote}
    Do you expect the National Health Service to get better, worse or stay the same over the next few years?
\end{quote}
We spell out the abbreviation `NHS' since we query for this survey question without additional context.
\revision{
We further extract public attitudes towards Boris Johnson as a Prime Minister with the following question~\cite{yougov_how_2024}:
\begin{quote}
    Do you think that Boris Johnson is doing well or badly as Prime Minister?
\end{quote}
And we evaluate public approval of the UK government with the following question~\cite{yougov_government_2024}:
\begin{quote}
    Do you approve or disapprove of the Government’s record to date?
\end{quote}
We include all respective answer options into our analysis and compare our results to survey data by \citeauthor{yougov_will_2024} for the respective question. Survey data is available in monthly waves for attitudes towards the NHS and towards Boris Johnson, and in weekly waves for Government Approval, with varying survey dates and varying numbers of participants.
}

\section{Temporal Adapters for LLMs}

Figure~\ref{fig:figure1} provides an overview of our proposed setup, comprising two separate steps: First, we fine-tune Temporal Adapters for Llama 3 8B on the Twitter panel data described in the previous section. Second, we prompt the model with one of the selected survey questions and one weekly adapter activated at a time. We extract token probabilities for each answer option across all weeks, which we combine into longitudinal macroscopes of emotions and attitudes.

To validate our results, we cross-correlate them with the respective survey data. We also include results from a BERT-based emotion detection model trained on labeled data~\cite{camacho-collados_TweetNLP_2022} for comparison. Finally, we conduct experiments on synthetically mixed, labeled data to demonstrate the internal validity of our attitude extraction method.

\subsection{Temporal Adapter Fine-Tuning}

Following the wording of YouGov's \textit{Britain's Mood, Measured Weekly}, we split our social media panel data into weekly subsets, containing seven days of text data leading up to the next YouGov survey wave. We obtain on average 385,000 tweets per training set, but the amount varies seasonally and around major events like the UK COVID-19 lockdown. We concatenate each training set into sequences of 512 tokens to facilitate batch training.

\begin{figure}[t!] 
    \centering
    \revfig{\includegraphics[width=0.9\columnwidth]{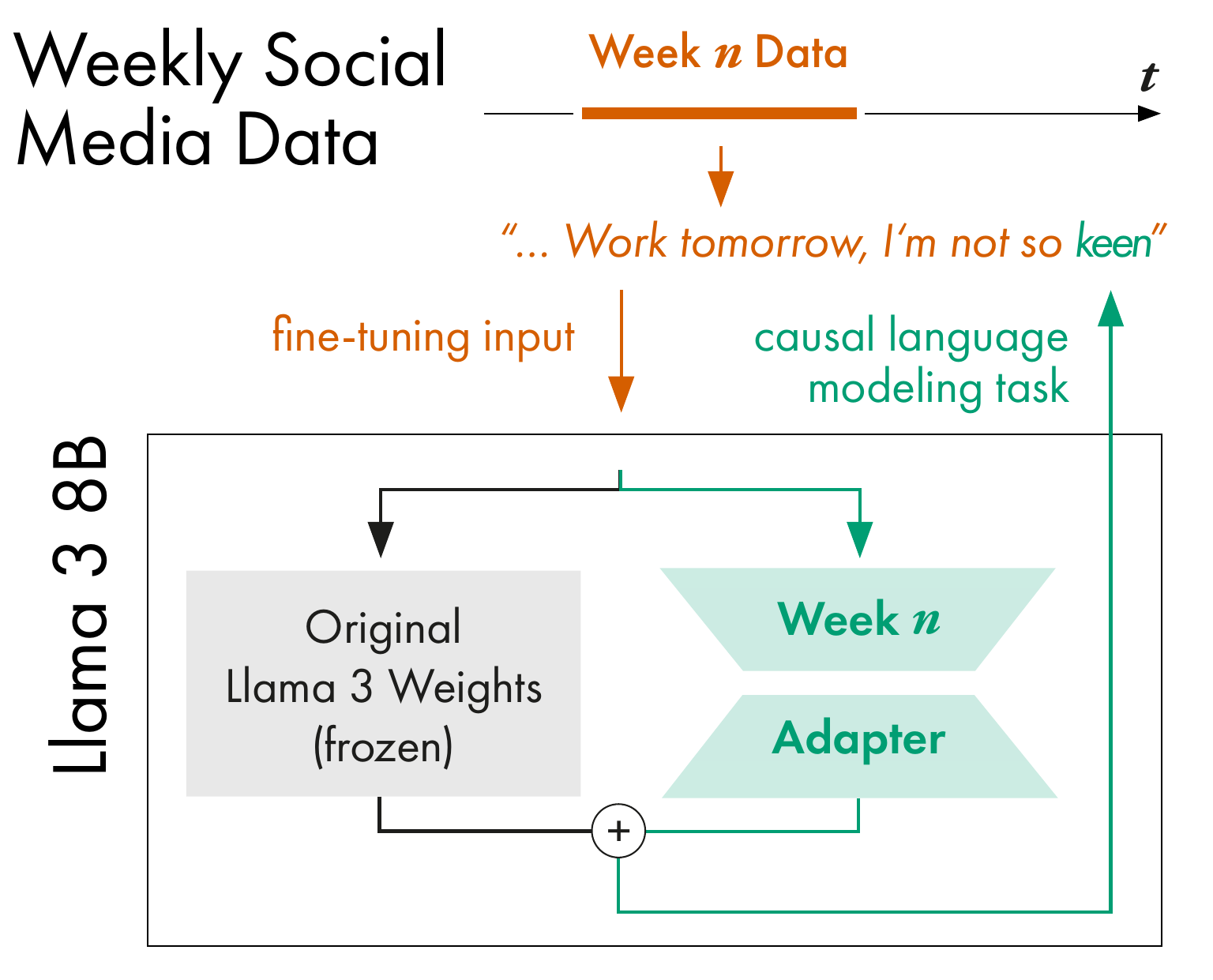}} 
    \caption{\textbf{Temporal Adapter Fine-Tuning.} We concatenate each week's tweets into chunks for batch-based fine-tuning. While fine-tuning each week's parameter-efficient LoRA~\cite{hu_lora_2021} adapter, the original model weights are kept frozen. Fine-Tuning is performed with the causal language modeling task, i.e., \revision{next-token prediction on the original tweets, independent of any survey data.}}
    \label{fig:lora_adapters}
\end{figure}

We select the base pretrained version (i.e., not instruction tuned) of a high-performing and openly available model, Llama 3 8B~\cite{aimeta_llama_2024}, to fine-tune the model on plain text data. We fine-tune \textit{Temporal Adapters} with the \textit{causal language modeling} objective, i.e. predicting next tokens, as shown in Figure~\ref{fig:lora_adapters}. \revision{We emphasize that the training procedure is completely independent of any survey data and only considers the word tokens that occur in the text data that we obtain from our Twitter panel. In other words, our Temporal Adapters fine-tune the LLM to more closely follow the token probabilities present in the tweets of a given week. We only consider survey data for evaluation, \textit{after} the training has been completed. Our method aims to fit the LLM to a specific week's tweets but the evaluation can only be performed across weeks after the training of all weeks has been completed. We thus opt to not have a test set of tweets in each week, as it would not prevent us from over-fitting to tweets from a specific week.}

Adapters are a parameter-efficient approach to fine-tuning LLMs that allows researchers to easily swap model properties~\cite{pfeiffer_modular_2023}. \revision{All types of adapters typically add a very small fraction of additional parameters to a language model with the goal to only train these parameters while keeping the original model weights frozen. This drastically improves training efficiency with little to no loss in training performance, as only relevant parts of the a model can be targeted during training. While many specific implementations of adapters have been proposed, LoRA adapters~\cite{hu_lora_2021} have evolved to be one of the most popular types.} They add rank-decomposition matrices to each layer of the transformer architecture. We train LoRA adapters of rank $128$ as a compromise between fine-tuning efficiency and training quality. This is twice the amount of parameters as the authors of the original LoRA paper considered necessary~\cite{hu_lora_2021}, but still only $0.67\%$ of the weights that would be trained when fully fine-tuning the model.

We conduct a partial hyperparameter search by training adapters with different learning rates, and for up to 8 epochs. We find that fine-tuning adapters works best for our purposes with a small learning rate of $5*10^{-6}$ and at most $1$ training epoch. The training loss is mostly stable after $0.5$ epochs at an average of $3.4$ across all adapters. In a preliminary experiment with a larger learning rate of $10^{-4}$, training loss decreased to around $2.6$ after $8$ training epochs but the cross-correlations of the attitudes we extracted with survey data were much lower. This is most likely due to the model over-fitting on the training data (tweets) that is quite different from the survey questions we prompt at inference time. \revision{We show performance of our method for varying LLM temperatures in Appendix Figure~\ref{fig:corr_by_temp}, and report results in the main paper for the Llama 3 default temperature if $0.6$.} We train adapters with \textit{3 different training seeds} for each week to better understand the reliability of our training method. We used \texttt{bfloat16} and the \texttt{adamw\_torch\_fused} optimizer with a batch size of $6$ and $4$ gradient accumulation steps. The training was conducted on two NVIDIA H100 GPUs in Distributed-Data-Parallel mode and takes approximately 20 minutes per adapter.

\subsection{Extracting Survey Answers}

Once we have finished adapter training, we focus on model inference and extract answers to survey questions as follows. We first concatenate a survey question with each of its $n$ answer options, obtaining $n$ separate prompts.
We experiment with an optional answer prefix that is build for the particular survey question. For instance, for \textit{Britain's Mood, Measured Weekly}, we add the optional answer prefix `I felt' which is followed by the answer options `happy', `sad' etc. This is to accommodate for the fact that first-token probabilities can be unreliable for scoring survey answers from LLMs~\cite{wang_my_2024}. We prompt each answer option separately and without additional labels as previous research has shown order effects and the tendency for LLMs to prefer certain answer labels~\cite{tjuatja_llms_2024, dominguez-olmedo_questioning_2023, wang_my_2024}.

\revision{We then extract answers from the token probabilities of the LLM as follows.} Based on a survey answer scoring method used in previous research~\cite{wu_towards_2024, naous_having_2024}, we perform a single forward pass of the concatenated question, optional answer prefix, and answer option through the LLM. We gather the token probabilities from the last LLM layer for each token in the answer option after applying softmax. If an answer option consists of multiple tokens, we multiply their probabilities.
Our extraction method is deterministic, i.e., if the weights and temperature of an LLM are kept constant, so will be the answer probabilities we extract. Unless otherwise noted, we report results for temperature \revision{$0.6$, i.e., the Llama 3 default.}

We perform survey answer extraction with each weekly adapter activated separately in the LLM and for each of the $3$ seeds. Since swapping adapters and inference is computationally inexpensive, we extract survey answers after training each adapter for $1$ epoch, as well as after every $50$ steps of training. We obtain a time series for each combination of question, answer options, and set of inference hyperparameters. \citeauthor{yougov_britains_2024} presents weekly estimates using a smoothed trend line to reduce random fluctuations due to sampling variability. We follow this approach and apply a $3$ week rolling average. For plotting time series, we apply min-max normalization and show the mean probability for each extracted survey answer across all $3$ seeds.

\subsection{\revision{Evaluation with Empirical Survey Data}}

\begin{figure*}[t!] 
    \centering
    \revfig{\includegraphics[width=\textwidth]{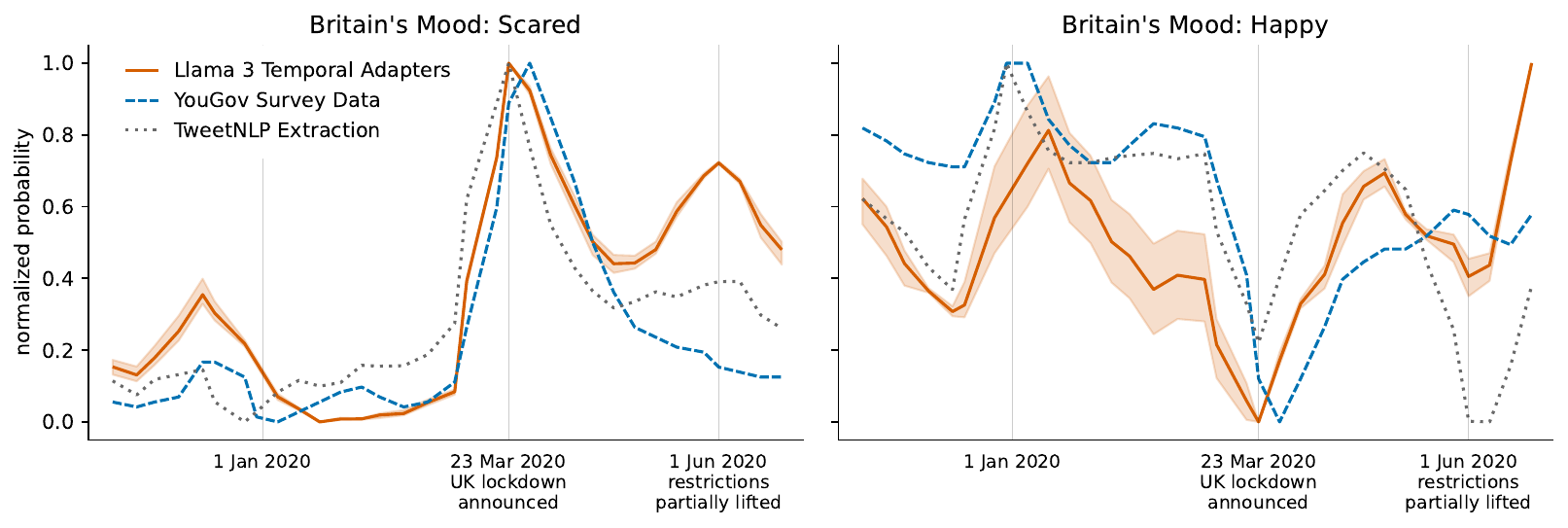}} 
    \caption{\textbf{Affect Aggregates Extracted from Temporal Adapters.} We extract answer probabilities by prompting a weekly fine-tuned Llama 3 8B with the same question wording as in the survey~\cite{yougov_britains_2024}, and compare them to the respective weekly survey data\revision{, as well as to a model that classifies selected emotions in Tweets directly (TweetNLP)}. The time series are min-max normalized and a $3$ week rolling average is applied. The shaded orange area indicates minimum and maximum LLM answer probabilities across $3$ training seeds. Our results descriptively show in the plot a similar trend of \revision{our estimate and the survey data} and we find strong positive and significant (\begingroup\setlength{\thickmuskip}{2mu}$p<0.01$\endgroup) cross-correlation between LLM probabilites and the survey data. Additional time series are provided in Figures~\ref{fig:fullTimeSeries_part1},~\ref{fig:fullTimeSeries_part2}, and\revision{~\ref{fig:fullTimeSeries_part3}} in the Appendix.
    }
    \label{fig:timeline_scared_sad}
\end{figure*}

\paragraph{Cross-Correlating Survey Data}
We evaluate our longitudinal macroscopes of emotions and attitudes by cross-correlating them with the respective survey data using Pearson's correlation coefficient. \citeauthor{yougov_britains_2024} publishes weekly survey results on \textit{Britain's Mood}, which results in $35$ data points in our observation period that we can use to compute cross-correlation. Results on British attitude towards the NHS \revision{and towards Boris Johnson as a Prime Minister} are only published monthly, i.e., we are left with 9 relevant data points.

We perform permutation tests with $10.000$ permutations\revision{, by shuffling the dates associated with each estimate we obtain from a specific Temporal Adapter, while keeping the survey data intact. This provides us with significance values for our cross-correlations and indicates for which estimates the longitudinal effect matters. In other words, significant results indicate that there are changes in the prevalence of an emotion/attitude over time and that we can successfully track these changes. Constant values or random noise in our estimates and the respective survey data would very unlikely produce significant results.}

\paragraph{Baseline Comparison Model}
We compare our results to emotion aggregates from a BERT-based model for emotion detection, TweetNLP~\cite{camacho-collados_TweetNLP_2022}. This baseline model will help us differentiate between issues in our training data, e.g. coverage error, or differences in expression of emotion, and measurement error specific to our research setup. TweetNLP was pre-trained on the \texttt{tweet\_eval} dataset that consists of tweets labeled with $11$ emotions~\cite{mohammad_semeval-2018_2018}. The model achieves a macro F1 score of $0.72$ on the \texttt{tweet\_eval} test set. We classify each tweet in our weekly training datasets separately and calculate the share of every emotion in every week. Being pre-trained on Twitter data makes TweetNLP particularly suited as a baseline model for our setup. However, only 4 of the emotions classified by TweetNLP match with emotions included in \textit{Britain's Mood, Measured Weekly}: `fear'/`scared', `sadness'/`sad', `joy'/`happy', and `optimism'/`optimistic'. 
\revision{This highlights the main issue of previous approaches that use classifiers like TweetNLP: They require expensively labeled training data or hand-crafted dictionaries. Our method, in contrast, does not require pre-defining a construct to be measured, but can be flexibly applied to any survey question.}
We again cross-correlate the results obtained from TweetNLP with the survey data and perform a permutation test.

\subsection{Synthetically Mixed Data}
To explore the internal validity of our emotion macroscopes, we perform additional experiments with synthetically mixed data that is labeled as `happy' or `sad'. With this setup, we can investigate whether a different prevalence of emotion signals indeed corresponds to different affect aggregates obtained from our method. We hypothesize a linear relationship between the amount of `sad' tweets in the training data and the token probability for `sad' as an answer option, and vice versa for `happy'. There are $1163$ tweets labeled `happy' and $1326$ tweets labeled `sad' in the \texttt{tweet\_eval} dataset. From these tweets, we select $1163$ using $11$ splits: $100 \%$ happy + $0 \%$ sad, $90 \%$ happy + $10 \%$ sad, ..., and $0 \%$ happy + $100 \%$ sad. We then train adapters on each of these splits using the same hyperparameters as described above. We repeat the splitting and training procedure using $10$ random seeds to improve robustness. Since the synthetically mixed training datasets are much smaller than our weekly Twitter datasets, the number of training steps/epochs is not directly comparable.

\section{Results}

\paragraph{Extracted Emotions Highly Correlate with Survey Data}
Figure~\ref{fig:timeline_scared_sad} shows time series for two collective emotions, scared and happy, that we extract using weekly Temporal Adapters. We use the original survey question developed by \citet{yougov_britains_2024} and compare our results with nationally representative survey data. We cap the training steps such that we have the same amount of training data across weeks.
The plot shows the mean answer probability across $3$ training seeds, with minimum and maximum probabilities, after a $3$ week rolling average was applied and after the time series were min-max normalized. Time series for other emotions are provided in Figures~\ref{fig:fullTimeSeries_part1} and~\ref{fig:fullTimeSeries_part2} in the Appendix.

Overall, we observe similar trends between our estimates of collective emotion and the survey data. \textit{Scared} spikes on the March 23\textsuperscript{rd}, 2020, when the first UK lockdown in response to the COVID-19 pandemic was announced.\footnote{\url{https://www.theguardian.com/world/2020/mar/23/boris-johnson-orders-uk-lockdown-to-be-enforced-by-police}, Accessed Sept 8\textsuperscript{th}, 2024} Another increase can be seen in our estimate of collective fear around June 1\textsuperscript{st} 2020, at a time when lockdown restrictions were partially lifted.\footnote{for England, see for instance \url{https://www.legislation.gov.uk/uksi/2020/558/made}, Accessed Sept 8\textsuperscript{th}, 2024} Fear also sees an increase around Christmas 2019. The survey data similarly indicates the largest level of collective fear around the announcement of the lockdown and a small increase around Christmas, but the second spike around June 1\textsuperscript{st}, 2020 is not observed in the survey. This is likely due to sampling errors in the social media data, i.e., more concerned people being active on Twitter, or due to differences in the expression of emotions on social media and in surveys. TweetNLP shows the same second spike in 2020 when used to classify the same training data, as shown in Figure~\ref{fig:timeline_scared_sad}.

For collective \textit{happiness}, we similarly extract estimates with good face validity. Happiness decreases rapidly in the two weeks before the nationwide lockdown was announced and recovers after another two weeks, presumably when people acclimatized with the new situation. This follows the trend that we observe in survey data. \revision{Similar to what we have observed for `scared'}, our estimate deviates from survey data around June 1\textsuperscript{st} 2020, when restrictions were partially lifted.
\revision{TweetNLP extracts the lowest happiness after June 1\textsuperscript{st}. Our method, instead, more closely tracks the \citeauthor{yougov_britains_2024} survey data and identifies the lowest happiness around March 23\textsuperscript{rd}.}
It should be addressed that we observe larger confidence intervals across $3$ training seeds, and larger week-by-week fluctuation for `happy' as compared to `scared'. One possible explanation for this would be that there is less information about happiness in our training data, which would lead to less influence on the fine-tuned adapters, and thus result in larger confidence intervals after min-max scaling. Another explanation could be that information about happiness is less evenly distributed in our training data. We sample an equal amount of text for each week and randomly shuffle the training data, so a skewed distribution of information about happiness would increase the relevance of different training seeds. Thirdly, \revision{`happy' could be a more difficult to learn concept for LLMs, potentially because it is expressed with more diverse language than `scared', leading to more error surrounding the concept in our outcome estimate}. 

\begin{figure}[t!] 
    \centering
    \revfig{\includegraphics[width=0.96\columnwidth]{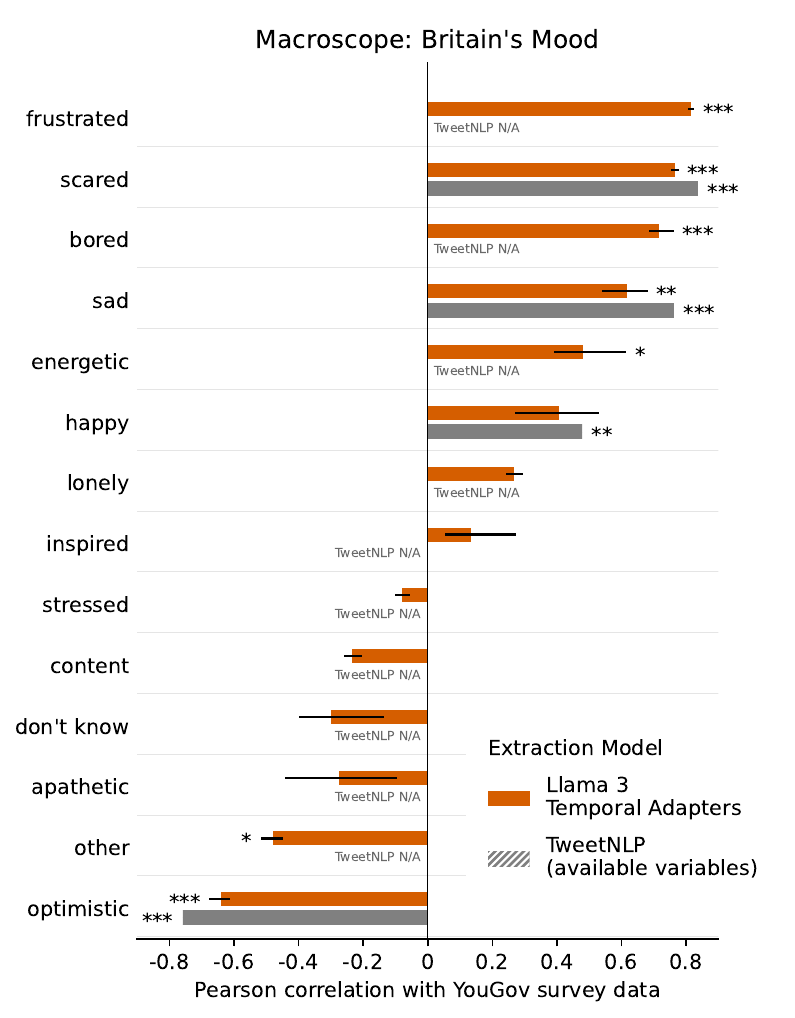}} 
    \caption{\textbf{Several Extracted Emotions Highly Correlate with Survey Data.} We cross-correlate the answers we extract from Llama 3 with the respective British survey data~\cite{yougov_britains_2024}.
    Across $3$ training seeds, we show minimum and maximum correlation with error bars and indicate the worst $p$ value
    \begingroup\setlength{\thickmuskip}{2mu}(*$p<0.05$,**$p<0.01$,***$p<0.001$)\endgroup.
    Our results vary strongly between emotions\revision{, but they are in line with TweetNLP's estimates. As opposed to TweetNLP, our method can be flexibly applied to extract any emotion as it does not require labeled training data.}
    }
    \label{fig:corr_mood}
\end{figure}

We calculate cross-correlations (Pearson) between the longitudinal data we extract and the respective survey data, and perform a significance test with $10,000$ permutations. Figure~\ref{fig:corr_mood} shows the results for all emotions found in \textit{Britain's Mood, Measured Weekly} across 3 training seeds, as well as the cross-correlation for our baseline method, TweetNLP~\cite{camacho-collados_TweetNLP_2022}. We observe three groups of emotions: (i) emotions that show a strong positive, significant correlation, (ii) emotions that are weakly positively or negatively correlated, and (iii) \textit{optimistic}, which is strongly negatively correlated. In the first group, we find both positively and negatively connoted emotions, and in particular affective phenomena that are sometimes referred to as being part of a set of \textit{basic emotions}: `happiness', `fear', and `sadness'~\cite{ekman_are_1992}. \revision{Significance levels obtained from a permutation test indicates that there are non-random changes in the emotions over time and that we can successfully track these changes.} For collective emotions where an estimate can be extracted from TweetNLP, i.e., where labeled training data was available, we find that our method produces comparable, \revision{yet slightly worse} results. \revision{We did not expect to outperform TweetNLP as it classifies individual tweets, while the Temporal Adapters in our setup have to aggregate information from many tweets first, before survey answers are predicted. We do, however,} obtain strong positive, significant correlations for emotions for which \textit{no} TweetNLP estimate is available: `frustrated', `bored', and `energetic'. \revision{This flexibility of our approach is a major benefit over existing methods such as TweetNLP and is further highlighted in the additional results reported below.} For group (ii)\revision{, emotions with weak positive/negative correlations}, it remains unclear why these collective emotions are hard to estimate from the data that we have available. One possible explanation might be that in contrast to the affective phenomena found in group (i), group (ii) contains more phenomena that clearly fall under `mood' rather than `emotion', i.e., they are more diffuse and longer lived, which might contradict typical social media behavior. Finally, we find a significant strong negative cross-correlation for `optimistic', both for our estimate as well as for the baseline model. This points to general differences in the expression of this emotions between surveys and social media.


\begin{table}[b!]
    \centering
    \small
    \setlength{\tabcolsep}{2mm}
    {\renewcommand{\arraystretch}{1.1}
    \revfig{\begin{tabular}{l|r|r|r}
                   & mean SE & median SE & SE std. dev.  \\ 
        \hline
        frustrated & 0.0414 & 0.0185 & 0.0507     \\
        scared     & 0.0509 & 0.0108 & 0.0872     \\
        bored      & 0.0387 & 0.0141 & 0.0586     \\
        sad        & 0.0502 & 0.0186 & 0.0666     \\
        energetic  & 0.0819 & 0.0531 & 0.0848     \\
        happy      & 0.0780 & 0.0482 & 0.0763     \\
        lonely     & 0.1229 & 0.0908 & 0.1256     \\
        inspired   & 0.1365 & 0.0737 & 0.1628     \\
        stressed   & 0.2284 & 0.0905 & 0.2865     \\
        content    & 0.1899 & 0.0806 & 0.2179     \\
        don't know & 0.1637 & 0.1439 & 0.1365     \\
        apathetic  & 0.2171 & 0.0987 & 0.2286     \\
        other      & 0.3119 & 0.3021 & 0.2347     \\
        optimistic & 0.3595 & 0.3598 & 0.2565    
    \end{tabular}
    }}
    \caption{\revision{
    \textbf{Means and Standard Deviations in Squared Errors (SE).}
    We individually min-max-normalize each time series -- survey data and Llama 3 estimates from 3 training seeds -- before we calculate squared errors between the same points in time in the YouGov survey data and the Llama 3 estimates. We find that strong positive cross-correlation shown in Figure~\ref{fig:corr_mood} coincides with low mean squared error.
    }}
    \label{tab:corr_mse}
\end{table}

\revision{For some collective emotions, we observe variable performance across time, i.e., our estimates obtained from Temporal Adapters closely track the YouGov survey data only for a certain time period. To address this issue, we additionally calculate squared errors between our estimates and the YouGov survey data, and report summary statistics of these squared errors in Table~\ref{tab:corr_mse}. Overall, we observe that strong positive cross-correlation for collective emotions such as \textit{frustrated, bored,} or \textit{scared} coincides with low mean squared errors and low standard deviations of these squared errors.}

\paragraph{Experiments with Synthetically Mixed Data}

\begin{figure}[t!] 
    \centering
    \includegraphics[width=0.6\columnwidth]{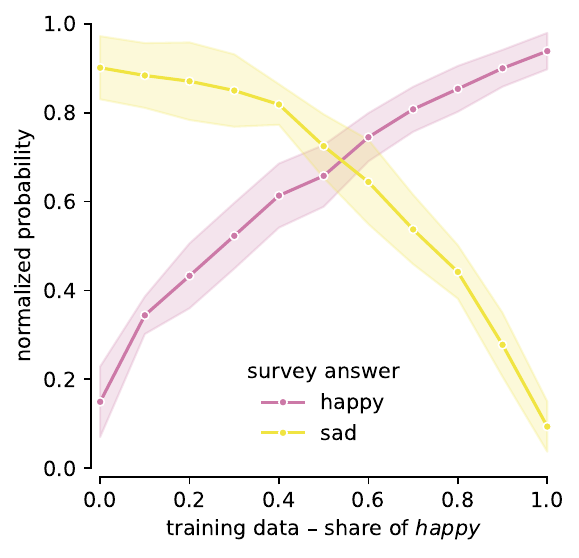} 
    \caption{\textbf{Internal Validity Demonstrated in Experiments with Synthetically Mixed Data.} We synthetically mix LLM training data with splits ranging from data that is labeled $100\%$ sad to $100\%$ happy. We then extract answers to \citet{yougov_britains_2024}'s survey question at each split, and
    show mean and standard deviation over 10 training seeds. The results support the internal validity of our extraction method, by showing a surprisingly linear relationship between training data ratio and extracted estimate, but also random training error.}
    \label{fig:synthMix}
\end{figure}

We create synthetically mixed training data from the \texttt{tweet\_eval} dataset~\cite{mohammad_semeval-2018_2018} in 11 splits ranging from $100\%$ sad to $100\%$ happy. We then prompt the model with the same question as used in \textit{Britain's Mood, Measured Weekly} and extract answer probabilities for the answer options `happy' and `sad'. We hypothesize a positive linear correlation between the share of `happy' tweets in the training data and the probability for answer option `happy', and vice versa for `sad'. Figure~\ref{fig:synthMix} shows the mean and standard deviation of extracted answers across 10 training seeds. We find strong positive correlations for both answer options, which supports the internal validity of our method -- more `happy' tweets actually lead to a higher answer probability for `happy' and vice versa for `sad'. For both emotions, the relationship between training data and extracted answer is surprisingly linear. Our answer scoring method does not ensure that the probabilities across all answer options add up to $1$, since we apply softmax on the last layer of the LLM, and each answer probability depends on probabilities of all other tokens in the vocabulary. The error bands indicate substantial random error resulting from the training and underline the importance of evaluating results obtained from multiple seeds.

In preliminary experiments, we investigated different hyperparameters for model training and for the extraction of affect aggregates, see Figure~\ref{fig:hyperparams} in the Appendix.
We investigated cross-correlations after different numbers of training steps and epochs and found little benefit in fine-tuning the model for more than $1$ epoch. While longer fine-tuning further improves training loss in the causal language modeling objective, the highest cross-correlations with survey data are observed relatively early on. This is likely due to the language model overfitting on the training data, which itself is quite different from the questions that we use for answer extraction. Similarly for learning rates, we found that smaller learning rates generally perform better, as long as loss does not remain constant. We find $5*10^{-6}$ to work best on our data. Finally, we found that training all adapters with the same amount of training data generally works better than training all adapters for $1$ epoch on all the data that was available in this week. In other words, we extracted answers for all Figures presented previously in this section after training each adapter for $350$ steps instead of $1$ epoch.

For answer extraction, we investigated the effect of temperature and of whether or not an answer prefix was attached to the answer options. A lower temperature leads to less evenly distributed token probabilities, similar to `enhancing contrast'. We found that this is beneficial when extracted attitudes have low noise, but that it can also increase random fluctuation. We also found that lower temperatures tend to overemphasize some answer options. Finally, adding an answer prefix leads to more consistent results. This is in line with previous findings on first-token probabilities~\cite{wang_my_2024} and is likely related to LLMs being trained to create text rather than to answer surveys with only an answer option.


\paragraph{Robustness Across Survey Instruments}

\begin{figure}[t!]
    \centering
    \begin{subfigure}{\columnwidth}
        \centering
        \caption{Britain's Mood: \textbf{Scared}}
        \revfig{\includegraphics[width=0.8\columnwidth]{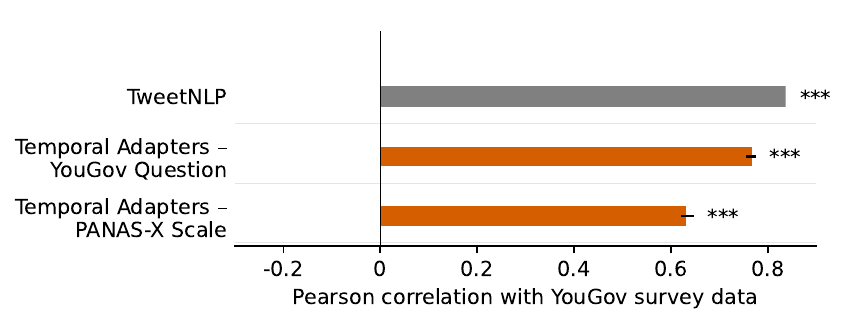}}
        \label{fig:corr_PANAS-X_scared}
    \end{subfigure}
    \begin{subfigure}{\columnwidth}
        \centering
        \caption{Britain's Mood: \textbf{Sad}}
        \revfig{\includegraphics[width=0.8\columnwidth]{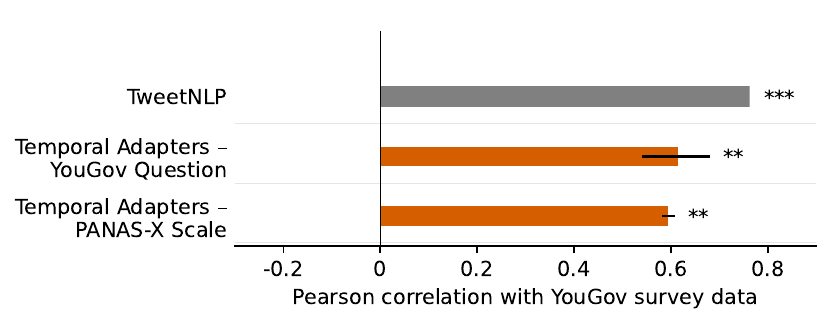}}
        \label{fig:corr_PANAS-X_sad}
    \end{subfigure}
    \caption{\revision{\textbf{Robustness across multiple survey instruments.} We extract the same collective emotions by querying Llama 3 with an additional survey instrument~\cite[PANAS-X:][]{clark_panas-x_1994} and cross-correlate our results with the respective \citeauthor{yougov_britains_2024} survey data.}
    Across $3$ training seeds, we indicate minimum and maximum correlation with error bars and show the worst $p$ value \begingroup\setlength{\thickmuskip}{2mu}(**$p<0.01$,***$p<0.001$)\endgroup. For time series, see Figure~\ref{fig:fullTimeSeries_PANAS-X} in the Appendix.
    }
    \label{fig:corr_PANAS-X}
\end{figure}

We implement two additional survey questions in addition to the one used in \textit{Britain's Mood, Measured Weekly}.
First, we demonstrate the robustness of our extraction method and extract the collective emotions \textit{scared} and \textit{sad} using the PANAS-X survey instrument. The instrument works quite differently from \citeauthor{yougov_britains_2024}'s questionnaire in that it measures each emotion with multiple adjectives, with the same answer options for each adjective: `very slightly or not at all', `a little', etc.
Again, we cross-correlate the extracted collective emotions with the \citeauthor{yougov_britains_2024} survey data -- Figure~\ref{fig:corr_PANAS-X} shows the results from 3 training seeds across the different extraction methods.
We find that both \textit{scared} and \textit{sad} achieve cross-correlations comparable to the ones we extracted with the original \citeauthor{yougov_britains_2024} question wording. This supports the robustness of our extraction method across prompt formulations, i.e., survey instruments, when measuring the same construct.

\begin{figure*}[ht!]
    \centering
    \begin{subfigure}{0.32\textwidth}
        \centering
        \caption{\textbf{The NHS will…}}
        \revfig{\includegraphics[width=\textwidth]{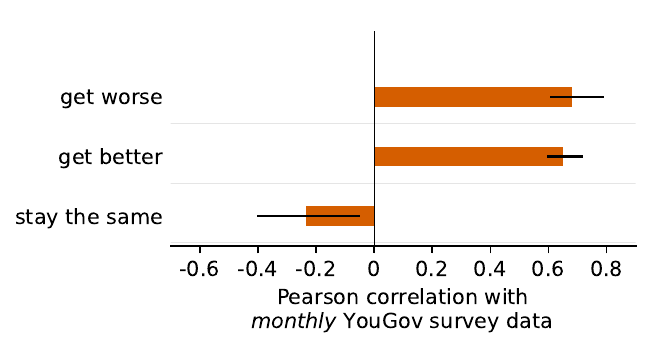}}
        \label{fig:corr_NHSbetter}
    \end{subfigure}
    \hfill
    \begin{subfigure}{0.32\textwidth}
        \centering
        \caption{\textbf{Boris Johnson is…}}
        \revfig{\includegraphics[width=\textwidth]{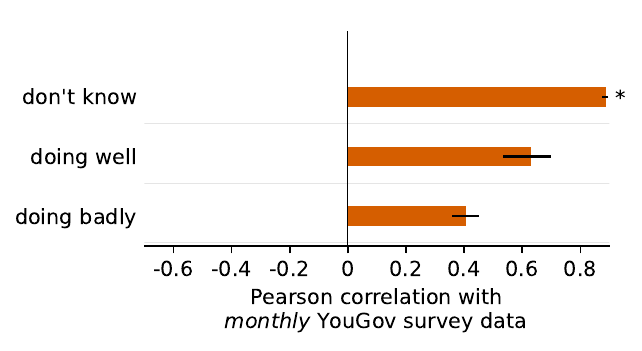}}
        \label{fig:corr_BorisPM}
    \end{subfigure}
    \hfill
    \begin{subfigure}{0.32\textwidth}
        \centering
        \caption{\textbf{Government Approval}}
        \revfig{\includegraphics[width=\textwidth]{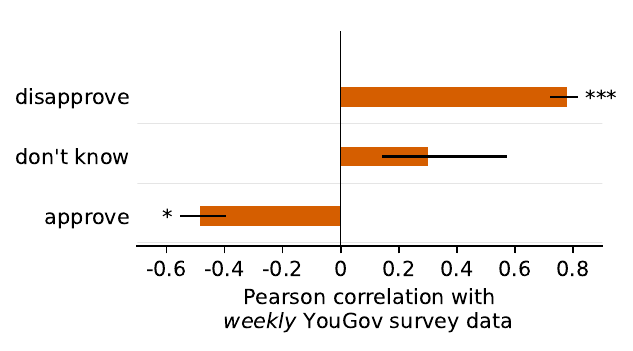}}
        \label{fig:corr_govApproval}
    \end{subfigure}
    \caption{\revision{\textbf{Successful Application to Additional Questions of Public Opinion Beyond Collective Emotions.} We extract collective attitudes towards the National Health Service (NHS), Boris Johnson as a Prime Minister, and the UK Government using survey questions developed by~\citet{yougov_will_2024, yougov_how_2024, yougov_government_2024}. We then cross-correlate our results with the respective YouGov survey data.
    Across $3$ training seeds, we indicate minimum and maximum correlation with error bars and show the worst $p$ value \begingroup\setlength{\thickmuskip}{2mu}(*$p<0.05$,***$p<0.001$)\endgroup. For time series, see Figures~\ref{fig:fullTimeSeries_NHSbetter}, \ref{fig:fullTimeSeries_BorisPM}, and~\ref{fig:fullTimeSeries_govApproval} in the Appendix. Our method can be used to extract collective answers to various issues of public opinion beyond collective emotions, as opposed to existing methods (such as TweetNLP), which have to be individually trained for specific questions.
    }}
    \label{fig:attitude_corr}
    \vspace{0.5cm}
\end{figure*}

\paragraph{Application to Attitude Aggregates}

Second, we go beyond measuring collective emotions and test a question that measures affective stance towards the National Health Service (NHS)\revision{, towards Boris Johnson as a Prime Minister, and towards the UK Government in general}. \citeauthor{yougov_will_2024} runs most of its `trackers' only in monthly waves, including on the attitude towards the NHS \revision{and towards Boris Johnson}. This means that during the time period that we study, we only have 9 data points available. This shows the benefits of extracting attitudes from social media data, an approach that is orders of magnitude cheaper as well as of higher temporal resolution than traditional surveys. Figure~\ref{fig:attitude_corr} shows the cross-correlation of the answer options for the 9 available data points between our estimates and the survey data -- for time series, see Figures~\ref{fig:fullTimeSeries_NHSbetter}, \revision{\ref{fig:fullTimeSeries_BorisPM}, and \ref{fig:fullTimeSeries_govApproval}} in the Appendix.
\revision{We find a strong positive cross-correlation between our estimate of government disapproval and the YouGov survey data. We also find moderate positive correlations for attitudes towards the NHS and towards Boris Johnson as a Prime Minister, given that we only have 9 data points available for evaluation. Existing NLP methods with pre-trained classifiers (such as TweetNLP) or dictionaries are always only suitable for extracting few, selected affect aggregates. This highlights a main benefit of our method compared to existing NLP methods, as it can be flexibly applied to answer any survey question without the need to re-train the Temporal Adapters.}



\section{Discussion}

\paragraph{\revision{Contribution}}
\revision{
This paper expands the inventory of methods for affect analysis available to our research community. We present a novel method for extracting longitudinal affect aggregates by using Temporal Adapters fine-tuned on user-generated data from social media. Our main focus lies on studying collective emotions in times of crisis, in particular, during the COVID-19 pandemic. Research on collective emotions has been shown to provide actionable insight for public health messaging and to study the social resilience of communities. We extend methods that have been previously used in this research area by proposing to extract survey answers with Temporal Adapters, rather than annotating and training specialized classification models. The main benefit of our approach is that is can be flexibly used with any survey question. This allows researchers to make use of the wealth of established survey questionnaires to extract affect aggregates from social media data, potentially from people who are inaccessible to traditional survey research. Ideally, such endeavors would use social media data from a clearly defined sample of users that represent the target population of a study. In the future, our method could allow for more timely estimates of affect aggregates and of public opinion, both within times of crisis and beyond that.
}

\revision{
Our work closes a temporal misalignment gap in previous surveys with LLMs by proposing Temporal Adapters that are trained on longitudinal social media data.
To the best of our knowledge, our paper is the first to extend the analysis of affect in LLMs to this \textit{longitudinal} setting.
Our approach creates novel ways to study affect aggregates in social media data longitudinally.
More broadly, they may also enable future, more temporally-aligned LLM-based studies of human attitudes, values, and opinions.
}

\paragraph{Limitations} Our method focuses on the extraction of separate time series data, rather than comparing the prevalence of emotions or attitudes cross-sectionally.
Our motivation behind that is that the relative probabilities of answer options in a cross-sectional sense solely depend on LLM pre-training.
Future research should combine these, so far rather disparate, approaches by correcting for differences that stem from model pre-training.
The results we present were obtained by training weekly Temporal Adapters, following exactly the intervals in which the respective survey waves were conducted.
This enables comparability between our estimates and survey data, but even higher temporal resolution than usually used in surveys, a potential benefit of using social media data, should be investigated in future research.
In contexts where survey data is available, survey data could in the future
be used as a prior for predicting affect aggregates with higher temporal resolution from social media data, i.e. filling the gaps in between waves. At the same, our method also offers a lot of potential in contexts in which no (recent) survey data is available, for diverse reasons such as lack of infrastructure, an ongoing violent conflict or similar factors leading to extreme non-response.
Like any other attitude extraction method, our approach requires validation in novel contexts, with possible design decisions ranging from model training hyperparameters and data sets used to the exact question wording and the answer extraction method.
\revision{Application to additional contexts and points in time is particularly important since we base our analysis on data collected during the unique situation of the COVID-19 pandemic.}
Future research should further investigate the applicability of this method to subgroup analysis, and in particular to whether affect aggregates
can be extracted with the same performance for different subpopulations, even if sampling error is overcome.
To facilitate such efforts, we publish our Python code under MIT license.\footnote{\revision{\url{https://github.com/dess-mannheim/temporal-adapters}}}

\paragraph{Ethical Considerations}

\citeauthor{yougov_panel_2024} pays its panel members adequately for their participation in surveys and is a member of several market research organisations (British Polling Council, ESOMAR, MRS) whose standards it adheres to~\cite{yougov_panel_2024}. We work with social media data that was publicly available at the time that we gathered it. We do not publish raw text data, or the Temporal Adapters that we trained, to protect the privacy of the individuals whose tweets are included in our training data. We do, however, publish the code that is needed to replicate our results under MIT license to support future research. Our method relies on more training data than the vast majority of social media users produce in a week, so it only works on an aggregate level. This limits the risks associated with profiling individuals through our method, even if we cannot fully rule out unethical applications. Our main results (Figure~\ref{fig:corr_mood}) highlight the importance of validation and comparison with survey data for each context and construct of interest. Still, an application of our method without such rigor could inform wrong or discriminatory downstream decisions, either because of bias in training data or because of measurement error.




\small
\bibliography{SurveyLLMs}

\begin{thebibliography}{62}
\providecommand{\natexlab}[1]{#1}

\bibitem[{Adilazuarda et~al.(2024)Adilazuarda, Mukherjee, Lavania, Singh, Dwivedi, Aji, O'Neill, Modi, and Choudhury}]{adilazuarda_towards_2024}
Adilazuarda, M.~F.; Mukherjee, S.; Lavania, P.; Singh, S.; Dwivedi, A.; Aji, A.~F.; O'Neill, J.; Modi, A.; and Choudhury, M. 2024.
\newblock Towards {Measuring} and {Modeling} "{Culture}" in {LLMs}: {A} {Survey}.
\newblock ArXiv:2403.15412 [cs].

\bibitem[{Agnew et~al.(2024)Agnew, Bergman, Chien, Díaz, El-Sayed, Pittman, Mohamed, and McKee}]{agnew_illusion_2024}
Agnew, W.; Bergman, A.~S.; Chien, J.; Díaz, M.; El-Sayed, S.; Pittman, J.; Mohamed, S.; and McKee, K.~R. 2024.
\newblock The illusion of artificial inclusion.
\newblock In \emph{Proceedings of the {CHI} {Conference} on {Human} {Factors} in {Computing} {Systems}}, 1--12.
\newblock ArXiv:2401.08572 [cs].

\bibitem[{AI@Meta(2024)}]{aimeta_llama_2024}
AI@Meta. 2024.
\newblock Llama 3 {Model} {Card}.
\newblock \url{https://github.com/meta-llama/llama3/blob/main/MODEL_CARD.md}.
\newblock Accessed: 2024-09-02.

\bibitem[{Amaya et~al.(2020)Amaya, Bach, Kreuter, and Keusch}]{hill_measuring_2020}
Amaya, A.; Bach, R.; Kreuter, F.; and Keusch, F. 2020.
\newblock Measuring the {Strength} of {Attitudes} in {Social} {Media} {Data}.
\newblock In Hill, C.~A.; Biemer, P.~P.; Buskirk, T.~D.; Japec, L.; Kirchner, A.; Kolenikov, S.; and Lyberg, L.~E., eds., \emph{Big {Data} {Meets} {Survey} {Science}}, 163--192. Wiley, 1 edition.
\newblock ISBN 978-1-118-97632-6 978-1-118-97635-7.

\bibitem[{Amba~Hombaiah et~al.(2021)Amba~Hombaiah, Chen, Zhang, Bendersky, and Najork}]{amba_hombaiah_dynamic_2021}
Amba~Hombaiah, S.; Chen, T.; Zhang, M.; Bendersky, M.; and Najork, M. 2021.
\newblock Dynamic {Language} {Models} for {Continuously} {Evolving} {Content}.
\newblock In \emph{Proceedings of the 27th {ACM} {SIGKDD} {Conference} on {Knowledge} {Discovery} \& {Data} {Mining}}, 2514--2524. Virtual Event Singapore: ACM.
\newblock ISBN 978-1-4503-8332-5.

\bibitem[{Argyle et~al.(2023)Argyle, Busby, Fulda, Gubler, Rytting, and Wingate}]{argyle_out_2023}
Argyle, L.~P.; Busby, E.~C.; Fulda, N.; Gubler, J.~R.; Rytting, C.; and Wingate, D. 2023.
\newblock Out of {One}, {Many}: {Using} {Language} {Models} to {Simulate} {Human} {Samples}.
\newblock \emph{Political Analysis}, 31(3): 337--351.

\bibitem[{Ashokkumar and Pennebaker(2021)}]{ashokkumar_social_2021}
Ashokkumar, A.; and Pennebaker, J.~W. 2021.
\newblock Social media conversations reveal large psychological shifts caused by {COVID}-19’s onset across {U}.{S}. cities.
\newblock \emph{Science Advances}, 7(39): eabg7843.

\bibitem[{Atari et~al.(2023)Atari, Xue, Park, Blasi, and Henrich}]{atari_which_2023}
Atari, M.; Xue, M.~J.; Park, P.~S.; Blasi, D.~E.; and Henrich, J. 2023.
\newblock Which {Humans}?
\newblock preprint, PsyArXiv.

\bibitem[{Bergman(1998)}]{bergman_theoretical_1998}
Bergman, M.~M. 1998.
\newblock A {Theoretical} {Note} on the {Differences} {Between} {Attitudes}, {Opinions}, and {Values}.
\newblock \emph{Swiss Political Science Review}, 4(2): 81--93.

\bibitem[{Bisbee et~al.(2023)Bisbee, Clinton, Dorff, Kenkel, and Larson}]{bisbee_synthetic_2023}
Bisbee, J.; Clinton, J.; Dorff, C.; Kenkel, B.; and Larson, J. 2023.
\newblock Synthetic {Replacements} for {Human} {Survey} {Data}? {The} {Perils} of {Large} {Language} {Models}.
\newblock preprint, SocArXiv.

\bibitem[{Brandwatch(2020{\natexlab{a}})}]{brandwatch_forsight_2020}
Brandwatch. 2020{\natexlab{a}}.
\newblock Forsight: {User} {Guide}.
\newblock \url{https://web.archive.org/web/20240513122948/https://www.brandwatch.com/wp-content/uploads/2020/10/Crimson-Hexagon-ForSight-User-Guide.pdf}.
\newblock Accessed: 2024-05-13.

\bibitem[{Brandwatch(2020{\natexlab{b}})}]{brandwatch_location_2020}
Brandwatch. 2020{\natexlab{b}}.
\newblock Location {Methodology}.
\newblock \url{https://web.archive.org/web/20210527091937/https://www.brandwatch.com/wp-content/uploads/2020/10/CrimsonHexagon_Location_Methodology.pdf}.
\newblock Accessed: 2021-05-27.

\bibitem[{Camacho-Collados et~al.(2022)Camacho-Collados, Rezaee, Riahi, Ushio, Loureiro, Antypas, Boisson, Espinosa-Anke, Liu, Martínez-Cámara, Medina, Buhrmann, Neves, and Barbieri}]{camacho-collados_TweetNLP_2022}
Camacho-Collados, J.; Rezaee, K.; Riahi, T.; Ushio, A.; Loureiro, D.; Antypas, D.; Boisson, J.; Espinosa-Anke, L.; Liu, F.; Martínez-Cámara, E.; Medina, G.; Buhrmann, T.; Neves, L.; and Barbieri, F. 2022.
\newblock {TweetNLP}: {Cutting}-{Edge} {Natural} {Language} {Processing} for {Social} {Media}.
\newblock ArXiv:2206.14774 [cs].

\bibitem[{Clark and Watson(1994)}]{clark_panas-x_1994}
Clark, L.~A.; and Watson, D. 1994.
\newblock The {PANAS}-{X}: {Manual} for the {Positive} and {Negative} {Affect} {Schedule} - {Expanded} {Form}.
\newblock Institution: University of Iowa.

\bibitem[{Dominguez-Olmedo, Hardt, and Mendler-Dünner(2023)}]{dominguez-olmedo_questioning_2023}
Dominguez-Olmedo, R.; Hardt, M.; and Mendler-Dünner, C. 2023.
\newblock Questioning the {Survey} {Responses} of {Large} {Language} {Models}.
\newblock ArXiv:2306.07951 [cs].

\bibitem[{Durmus et~al.(2023)Durmus, Nyugen, Liao, Schiefer, Askell, Bakhtin, Chen, Hatfield-Dodds, Hernandez, Joseph, Lovitt, McCandlish, Sikder, Tamkin, Thamkul, Kaplan, Clark, and Ganguli}]{durmus_towards_2023}
Durmus, E.; Nyugen, K.; Liao, T.~I.; Schiefer, N.; Askell, A.; Bakhtin, A.; Chen, C.; Hatfield-Dodds, Z.; Hernandez, D.; Joseph, N.; Lovitt, L.; McCandlish, S.; Sikder, O.; Tamkin, A.; Thamkul, J.; Kaplan, J.; Clark, J.; and Ganguli, D. 2023.
\newblock Towards {Measuring} the {Representation} of {Subjective} {Global} {Opinions} in {Language} {Models}.
\newblock ArXiv:2306.16388 [cs].

\bibitem[{Ekman(1992)}]{ekman_are_1992}
Ekman, P. 1992.
\newblock Are there basic emotions?
\newblock \emph{Psychological Review}, 99(3): 550--553.

\bibitem[{FORCE11(2020)}]{force11_fair_2020}
FORCE11. 2020.
\newblock The {FAIR} {Data} principles.
\newblock \url{https://force11.org/info/the-fair-data-principles/}.

\bibitem[{Garcia et~al.(2021)Garcia, Pellert, Lasser, and Metzler}]{garcia_social_2021}
Garcia, D.; Pellert, M.; Lasser, J.; and Metzler, H. 2021.
\newblock Social media emotion macroscopes reflect emotional experiences in society at large.
\newblock ArXiv:2107.13236 [cs].

\bibitem[{Garcia and Rimé(2019)}]{garcia_collective_2019}
Garcia, D.; and Rimé, B. 2019.
\newblock Collective {Emotions} and {Social} {Resilience} in the {Digital} {Traces} {After} a {Terrorist} {Attack}.
\newblock \emph{Psychological Science}, 30(4): 617--628.

\bibitem[{Gayo-Avello(2013)}]{gayo-avello_meta-analysis_2013}
Gayo-Avello, D. 2013.
\newblock A {Meta}-{Analysis} of {State}-of-the-{Art} {Electoral} {Prediction} {From} {Twitter} {Data}.
\newblock \emph{Social Science Computer Review}, 31(6): 649--679.

\bibitem[{Gebru et~al.(2021)Gebru, Morgenstern, Vecchione, Vaughan, Wallach, Iii, and Crawford}]{gebru_datasheets_2021}
Gebru, T.; Morgenstern, J.; Vecchione, B.; Vaughan, J.~W.; Wallach, H.; Iii, H.~D.; and Crawford, K. 2021.
\newblock Datasheets for datasets.
\newblock \emph{Communications of the ACM}, 64(12): 86--92.

\bibitem[{Goldenberg et~al.(2020)Goldenberg, Garcia, Halperin, and Gross}]{goldenberg_collective_2020}
Goldenberg, A.; Garcia, D.; Halperin, E.; and Gross, J.~J. 2020.
\newblock Collective {Emotions}.
\newblock \emph{Current Directions in Psychological Science}, 29(2): 154--160.

\bibitem[{Golder and Macy(2011)}]{golder_diurnal_2011}
Golder, S.~A.; and Macy, M.~W. 2011.
\newblock Diurnal and {Seasonal} {Mood} {Vary} with {Work}, {Sleep}, and {Daylength} {Across} {Diverse} {Cultures}.
\newblock \emph{Science}, 333(6051): 1878--1881.

\bibitem[{Hartmann, Schwenzow, and Witte(2023)}]{hartmann_political_2023}
Hartmann, J.; Schwenzow, J.; and Witte, M. 2023.
\newblock The political ideology of conversational {AI}: {Converging} evidence on {ChatGPT}’s pro-environmental, left-libertarian orientation.

\bibitem[{Hu et~al.(2021)Hu, Shen, Wallis, Allen-Zhu, Li, Wang, Wang, and Chen}]{hu_lora_2021}
Hu, E.~J.; Shen, Y.; Wallis, P.; Allen-Zhu, Z.; Li, Y.; Wang, S.; Wang, L.; and Chen, W. 2021.
\newblock {LoRA}: {Low}-{Rank} {Adaptation} of {Large} {Language} {Models}.
\newblock ArXiv:2106.09685 [cs].

\bibitem[{Jones and Silver(2020)}]{jones_this_2020}
Jones, N.~M.; and Silver, R.~C. 2020.
\newblock This is not a drill: {Anxiety} on {Twitter} following the 2018 {Hawaii} false missile alert.
\newblock \emph{American Psychologist}, 75(5): 683--693.

\bibitem[{Kim and Lee(2023)}]{kim_ai-augmented_2023}
Kim, J.; and Lee, B. 2023.
\newblock {AI}-{Augmented} {Surveys}: {Leveraging} {Large} {Language} {Models} and {Surveys} for {Opinion} {Prediction}.
\newblock ArXiv:2305.09620 [cs].

\bibitem[{Lazaridou et~al.(2021)Lazaridou, Kuncoro, Gribovskaya, Agrawal, Liška, Terzi, and Gimenez}]{lazaridou_mind_2021}
Lazaridou, A.; Kuncoro, A.; Gribovskaya, E.; Agrawal, D.; Liška, A.; Terzi, T.; and Gimenez, M. 2021.
\newblock Mind the {Gap}: {Assessing} {Temporal} {Generalization} in {Neural} {Language} {Models}.
\newblock \emph{Advances in Neural Information Processing Systems}, 34: 29348--29363.

\bibitem[{Lively and Weed(2016)}]{lively_sociology_2016}
Lively, K.~J.; and Weed, E.~A. 2016.
\newblock The {Sociology} of {Emotions}.
\newblock In Feldman~Barret, L.; Lewis, M.; and Haviland-Jones, J.~M., eds., \emph{Handbook of {Emotions}}, 66--81. New York: Guilford Press, fourth edition.

\bibitem[{Lwin et~al.(2020)Lwin, Lu, Sheldenkar, Schulz, Shin, Gupta, and Yang}]{lwin_global_2020}
Lwin, M.~O.; Lu, J.; Sheldenkar, A.; Schulz, P.~J.; Shin, W.; Gupta, R.; and Yang, Y. 2020.
\newblock Global {Sentiments} {Surrounding} the {COVID}-19 {Pandemic} on {Twitter}: {Analysis} of {Twitter} {Trends}.
\newblock \emph{JMIR Public Health and Surveillance}, 6(2): e19447.

\bibitem[{Ma et~al.(2024)Ma, Wang, Hu, Haensch, Hedderich, Plank, and Kreuter}]{ma_potential_2024}
Ma, B.; Wang, X.; Hu, T.; Haensch, A.-C.; Hedderich, M.~A.; Plank, B.; and Kreuter, F. 2024.
\newblock The {Potential} and {Challenges} of {Evaluating} {Attitudes}, {Opinions}, and {Values} in {Large} {Language} {Models}.
\newblock ArXiv:2406.11096 [cs].

\bibitem[{McIlroy-Young et~al.(2024)McIlroy-Young, Brown, Olson, Zhang, and Dwork}]{mcilroy-young_set-based_2024}
McIlroy-Young, R.; Brown, K.; Olson, C.; Zhang, L.; and Dwork, C. 2024.
\newblock Set-{Based} {Prompting}: {Provably} {Solving} the {Language} {Model} {Order} {Dependency} {Problem}.
\newblock ArXiv:2406.06581 [cs].

\bibitem[{Metzler et~al.(2023)Metzler, Rimé, Pellert, Niederkrotenthaler, Di~Natale, and Garcia}]{metzler_collective_2023}
Metzler, H.; Rimé, B.; Pellert, M.; Niederkrotenthaler, T.; Di~Natale, A.; and Garcia, D. 2023.
\newblock Collective emotions during the {COVID}-19 outbreak.
\newblock \emph{Emotion}, 23(3): 844--858.

\bibitem[{Mohammad et~al.(2018)Mohammad, Bravo-Marquez, Salameh, and Kiritchenko}]{mohammad_semeval-2018_2018}
Mohammad, S.; Bravo-Marquez, F.; Salameh, M.; and Kiritchenko, S. 2018.
\newblock {SemEval}-2018 {Task} 1: {Affect} in {Tweets}.
\newblock In \emph{Proceedings of {The} 12th {International} {Workshop} on {Semantic} {Evaluation}}, 1--17. New Orleans, Louisiana: Association for Computational Linguistics.

\bibitem[{Mohiyeddini and Bauer(2013)}]{mohiyeddini_what_2013}
Mohiyeddini, C.; and Bauer, S. 2013.
\newblock What is an {Emotion}?
\newblock In Mohiyeddini, C.; Eysenck, M.~W.; and Bauer, S., eds., \emph{Handbook of {Psychology} of {Emotions}}, volume Volume 1: Recent Theoretical Perspectives and Novel Empirical Findings, 3--10. New York: Nova Publ.
\newblock ISBN 978-1-62808-053-7.

\bibitem[{Motoki, Pinho~Neto, and Rodrigues(2023)}]{motoki_more_2023}
Motoki, F.; Pinho~Neto, V.; and Rodrigues, V. 2023.
\newblock More human than human: measuring {ChatGPT} political bias.
\newblock \emph{Public Choice}, 198: 3--23.

\bibitem[{Naous et~al.(2024)Naous, Ryan, Ritter, and Xu}]{naous_having_2024}
Naous, T.; Ryan, M.~J.; Ritter, A.; and Xu, W. 2024.
\newblock Having {Beer} after {Prayer}? {Measuring} {Cultural} {Bias} in {Large} {Language} {Models}.
\newblock ArXiv:2305.14456 [cs].

\bibitem[{Oatley and Johnson-Laird(2014)}]{oatley_cognitive_2014}
Oatley, K.; and Johnson-Laird, P. 2014.
\newblock Cognitive approaches to emotions.
\newblock \emph{Trends in Cognitive Sciences}, 18(3): 134--140.

\bibitem[{Olteanu et~al.(2019)Olteanu, Castillo, Diaz, and Kıcıman}]{olteanu_social_2019}
Olteanu, A.; Castillo, C.; Diaz, F.; and Kıcıman, E. 2019.
\newblock Social {Data}: {Biases}, {Methodological} {Pitfalls}, and {Ethical} {Boundaries}.
\newblock \emph{Frontiers in Big Data}, 2: 13.

\bibitem[{Pasek et~al.(2018)Pasek, Yan, Conrad, Newport, and Marken}]{pasek_stability_2018}
Pasek, J.; Yan, H.~Y.; Conrad, F.~G.; Newport, F.; and Marken, S. 2018.
\newblock The {Stability} of {Economic} {Correlations} over {Time}.
\newblock \emph{Public Opinion Quarterly}, 82(3): 470--492.

\bibitem[{Pellert et~al.(2022)Pellert, Metzler, Matzenberger, and Garcia}]{pellert_validating_2022}
Pellert, M.; Metzler, H.; Matzenberger, M.; and Garcia, D. 2022.
\newblock Validating daily social media macroscopes of emotions.
\newblock \emph{Scientific Reports}, 12(1): 11236.

\bibitem[{Pfeiffer et~al.(2023)Pfeiffer, Ruder, Vulić, and Ponti}]{pfeiffer_modular_2023}
Pfeiffer, J.; Ruder, S.; Vulić, I.; and Ponti, E.~M. 2023.
\newblock Modular {Deep} {Learning}.
\newblock ArXiv:2302.11529 [cs].

\bibitem[{Ramezani and Xu(2023)}]{ramezani_knowledge_2023}
Ramezani, A.; and Xu, Y. 2023.
\newblock Knowledge of cultural moral norms in large language models.
\newblock ArXiv:2306.01857 [cs].

\bibitem[{Rogers and Robinson(2014)}]{stets_measuring_2014}
Rogers, K.~B.; and Robinson, D.~T. 2014.
\newblock Measuring {Affect} and {Emotions}.
\newblock In Stets, J.~E.; and Turner, J.~H., eds., \emph{Handbook of the {Sociology} of {Emotions}: {Volume} {II}}, 283--303. Dordrecht: Springer Netherlands.
\newblock ISBN 978-94-017-9129-8 978-94-017-9130-4.
\newblock Series Title: Handbooks of Sociology and Social Research.

\bibitem[{Röttger et~al.(2024)Röttger, Hofmann, Pyatkin, Hinck, Kirk, Schütze, and Hovy}]{rottger_political_2024}
Röttger, P.; Hofmann, V.; Pyatkin, V.; Hinck, M.; Kirk, H.~R.; Schütze, H.; and Hovy, D. 2024.
\newblock Political {Compass} or {Spinning} {Arrow}? {Towards} {More} {Meaningful} {Evaluations} for {Values} and {Opinions} in {Large} {Language} {Models}.
\newblock ArXiv:2402.16786 [cs].

\bibitem[{Röttger and Pierrehumbert(2021)}]{rottger_temporal_2021}
Röttger, P.; and Pierrehumbert, J.~B. 2021.
\newblock Temporal {Adaptation} of {BERT} and {Performance} on {Downstream} {Document} {Classification}: {Insights} from {Social} {Media}.
\newblock ArXiv:2104.08116 [cs].

\bibitem[{Santurkar et~al.(2023)Santurkar, Durmus, Ladhak, Lee, Liang, and Hashimoto}]{santurkar_whose_2023}
Santurkar, S.; Durmus, E.; Ladhak, F.; Lee, C.; Liang, P.; and Hashimoto, T. 2023.
\newblock Whose {Opinions} {Do} {Language} {Models} {Reflect}?
\newblock ArXiv:2303.17548 [cs].

\bibitem[{Sen et~al.(2021)Sen, Flöck, Weller, Weiß, and Wagner}]{sen_total_2021}
Sen, I.; Flöck, F.; Weller, K.; Weiß, B.; and Wagner, C. 2021.
\newblock A {Total} {Error} {Framework} for {Digital} {Traces} of {Human} {Behavior} on {Online} {Platforms}.
\newblock \emph{Public Opinion Quarterly}, 85(S1): 399--422.

\bibitem[{Sun et~al.(2024)Sun, Lee, Nan, Zhao, Lee, Jansen, and Kim}]{sun_random_2024}
Sun, S.; Lee, E.; Nan, D.; Zhao, X.; Lee, W.; Jansen, B.~J.; and Kim, J.~H. 2024.
\newblock Random {Silicon} {Sampling}: {Simulating} {Human} {Sub}-{Population} {Opinion} {Using} a {Large} {Language} {Model} {Based} on {Group}-{Level} {Demographic} {Information}.
\newblock Version Number: 1.

\bibitem[{Tjuatja et~al.(2024)Tjuatja, Chen, Wu, Talwalkar, and Neubig}]{tjuatja_llms_2024}
Tjuatja, L.; Chen, V.; Wu, S.~T.; Talwalkar, A.; and Neubig, G. 2024.
\newblock Do {LLMs} exhibit human-like response biases? {A} case study in survey design.
\newblock ArXiv:2311.04076 [cs].

\bibitem[{Valdez et~al.(2020)Valdez, Ten~Thij, Bathina, Rutter, and Bollen}]{valdez_social_2020}
Valdez, D.; Ten~Thij, M.; Bathina, K.; Rutter, L.~A.; and Bollen, J. 2020.
\newblock Social {Media} {Insights} {Into} {US} {Mental} {Health} {During} the {COVID}-19 {Pandemic}: {Longitudinal} {Analysis} of {Twitter} {Data}.
\newblock \emph{Journal of Medical Internet Research}, 22(12): e21418.

\bibitem[{Von Der~Heyde, Haensch, and Wenz(2023)}]{von_der_heyde_vox_2023}
Von Der~Heyde, L.; Haensch, A.-C.; and Wenz, A. 2023.
\newblock Vox {Populi}, {Vox} {AI}? {Using} {Language} {Models} to {Estimate} {German} {Public} {Opinion}.
\newblock preprint, SocArXiv.

\bibitem[{Wang, Morgenstern, and Dickerson(2024)}]{wang_large_2024}
Wang, A.; Morgenstern, J.; and Dickerson, J.~P. 2024.
\newblock Large language models cannot replace human participants because they cannot portray identity groups.
\newblock ArXiv:2402.01908 [cs].

\bibitem[{Wang et~al.(2024{\natexlab{a}})Wang, Hu, Ma, Röttger, and Plank}]{wang_look_2024}
Wang, X.; Hu, C.; Ma, B.; Röttger, P.; and Plank, B. 2024{\natexlab{a}}.
\newblock Look at the {Text}: {Instruction}-{Tuned} {Language} {Models} are {More} {Robust} {Multiple} {Choice} {Selectors} than {You} {Think}.
\newblock ArXiv:2404.08382 [cs].

\bibitem[{Wang et~al.(2024{\natexlab{b}})Wang, Ma, Hu, Weber-Genzel, Röttger, Kreuter, Hovy, and Plank}]{wang_my_2024}
Wang, X.; Ma, B.; Hu, C.; Weber-Genzel, L.; Röttger, P.; Kreuter, F.; Hovy, D.; and Plank, B. 2024{\natexlab{b}}.
\newblock "{My} {Answer} is {C}": {First}-{Token} {Probabilities} {Do} {Not} {Match} {Text} {Answers} in {Instruction}-{Tuned} {Language} {Models}.
\newblock ArXiv:2402.14499 [cs].

\bibitem[{Wu et~al.(2024)Wu, Khan, Das, Nanda, Ghosh, Kolling, Speicher, Bindschaedler, Gummadi, and Terzi}]{wu_towards_2024}
Wu, Q.; Khan, M.~A.; Das, S.; Nanda, V.; Ghosh, B.; Kolling, C.; Speicher, T.; Bindschaedler, L.; Gummadi, K.~P.; and Terzi, E. 2024.
\newblock Towards {Reliable} {Latent} {Knowledge} {Estimation} in {LLMs}: {In}-{Context} {Learning} vs. {Prompting} {Based} {Factual} {Knowledge} {Extraction}.
\newblock ArXiv:2404.12957 [cs].

\bibitem[{YouGov(2024{\natexlab{a}})}]{yougov_britains_2024}
YouGov. 2024{\natexlab{a}}.
\newblock Britain's {Mood}, {Measured} {Weekly}.
\newblock \url{https://yougov.co.uk/topics/politics/trackers/britains-mood-measured-weekly}.
\newblock Accessed: 2024-09-02.

\bibitem[{YouGov(2024{\natexlab{b}})}]{yougov_government_2024}
YouGov. 2024{\natexlab{b}}.
\newblock Government approval.
\newblock \url{https://yougov.co.uk/topics/politics/trackers/government-approval}.
\newblock Accessed: 2024-12-30.

\bibitem[{YouGov(2024{\natexlab{c}})}]{yougov_how_2024}
YouGov. 2024{\natexlab{c}}.
\newblock How well is {Boris} {Johnson} doing as {Prime} {Minister}?
\newblock \url{https://yougov.co.uk/topics/politics/trackers/boris-johnson-approval-rating}.
\newblock Accessed: 2024-12-30.

\bibitem[{YouGov(2024{\natexlab{d}})}]{yougov_panel_2024}
YouGov. 2024{\natexlab{d}}.
\newblock Panel {Methodology}.
\newblock \url{https://yougov.co.uk/about/panel-methodology}.
\newblock Accessed: 2024-09-02.

\bibitem[{YouGov(2024{\natexlab{e}})}]{yougov_will_2024}
YouGov. 2024{\natexlab{e}}.
\newblock Will the {NHS} get better or worse?
\newblock \url{https://yougov.co.uk/topics/politics/trackers/is-the-nhs-getting-better-or-worse}.
\newblock Accessed: 2024-09-02.

\end{thebibliography}

\section*{Paper Checklist}

\begin{enumerate}

\item For most authors...
\begin{enumerate}
    \item  Would answering this research question advance science without violating social contracts, such as violating privacy norms, perpetuating unfair profiling, exacerbating the socio-economic divide, or implying disrespect to societies or cultures?
    \answerYes{Yes, see Ethical Considerations}
  \item Do your main claims in the abstract and introduction accurately reflect the paper's contributions and scope?
    \answerYes{Yes}
   \item Do you clarify how the proposed methodological approach is appropriate for the claims made? 
    \answerYes{Yes, see Comparison and Validation}
   \item Do you clarify what are possible artifacts in the data used, given population-specific distributions?
    \answerYes{Yes, see Social Media Macroscopes}
  \item Did you describe the limitations of your work?
    \answerYes{Yes, see Limitations}
  \item Did you discuss any potential negative societal impacts of your work?
    \answerYes{Yes, see Ethical Considerations}
      \item Did you discuss any potential misuse of your work?
    \answerYes{Yes, see Ethical Considerations}
    \item Did you describe steps taken to prevent or mitigate potential negative outcomes of the research, such as data and model documentation, data anonymization, responsible release, access control, and the reproducibility of findings?
    \answerYes{Yes, see Ethical Considerations}
  \item Have you read the ethics review guidelines and ensured that your paper conforms to them?
    \answerYes{Yes}
\end{enumerate}

\item Additionally, if your study involves hypotheses testing...
\begin{enumerate}
  \item Did you clearly state the assumptions underlying all theoretical results?
    \answerNA{NA}
  \item Have you provided justifications for all theoretical results?
    \answerNA{NA}
  \item Did you discuss competing hypotheses or theories that might challenge or complement your theoretical results?
    \answerNA{NA}
  \item Have you considered alternative mechanisms or explanations that might account for the same outcomes observed in your study?
    \answerNA{NA}
  \item Did you address potential biases or limitations in your theoretical framework?
    \answerNA{NA}
  \item Have you related your theoretical results to the existing literature in social science?
    \answerNA{NA}
  \item Did you discuss the implications of your theoretical results for policy, practice, or further research in the social science domain?
    \answerNA{NA}
\end{enumerate}

\item Additionally, if you are including theoretical proofs...
\begin{enumerate}
  \item Did you state the full set of assumptions of all theoretical results?
    \answerNA{NA}
	\item Did you include complete proofs of all theoretical results?
    \answerNA{NA}
\end{enumerate}

\item Additionally, if you ran machine learning experiments...
\begin{enumerate}
  \item Did you include the code, data, and instructions needed to reproduce the main experimental results (either in the supplemental material or as a URL)?
    \answerYes{Yes, we publish the code under MIT license}
  \item Did you specify all the training details (e.g., data splits, hyperparameters, how they were chosen)?
    \answerYes{Yes}
     \item Did you report error bars (e.g., with respect to the random seed after running experiments multiple times)?
    \answerYes{Yes}
	\item Did you include the total amount of compute and the type of resources used (e.g., type of GPUs, internal cluster, or cloud provider)?
    \answerYes{Yes}
     \item Do you justify how the proposed evaluation is sufficient and appropriate to the claims made? 
    \answerYes{Yes, see Comparison and Validation}
     \item Do you discuss what is ``the cost`` of misclassification and fault (in)tolerance?
    \answerYes{Yes, see Ethical Considerations}
  
\end{enumerate}

\item Additionally, if you are using existing assets (e.g., code, data, models) or curating/releasing new assets, \textbf{without compromising anonymity}...
\begin{enumerate}
  \item If your work uses existing assets, did you cite the creators?
    \answerYes{Yes, see Longitudinal Datasets}
  \item Did you mention the license of the assets?
    \answerYes{Yes}
  \item Did you include any new assets in the supplemental material or as a URL?
    \answerNo{No, to protect individual privacy}
  \item Did you discuss whether and how consent was obtained from people whose data you're using/curating?
    \answerYes{Yes, see Ethical Considerations}
  \item Did you discuss whether the data you are using/curating contains personally identifiable information or offensive content?
    \answerYes{Yes, see Ethical Considerations}
  \item If you are curating or releasing new datasets, did you discuss how you intend to make your datasets FAIR (see \citet{force11_fair_2020})?
    \answerNA{NA}
  \item If you are curating or releasing new datasets, did you create a Datasheet for the Dataset (see \citet{gebru_datasheets_2021})? 
    \answerNA{NA}
\end{enumerate}

\item Additionally, if you used crowdsourcing or conducted research with human subjects, \textbf{without compromising anonymity}...
\begin{enumerate}
  \item Did you include the full text of instructions given to participants and screenshots?
    \answerYes{Yes, see Questionnaires and Survey Data}
  \item Did you describe any potential participant risks, with mentions of Institutional Review Board (IRB) approvals?
    \answerNo{No, because survey data was gathered by YouGov}
  \item Did you include the estimated hourly wage paid to participants and the total amount spent on participant compensation?
    \answerNo{No. YouGov does not publish detailed enough information, but adheres to ESOMAR/MRS standards.}
   \item Did you discuss how data is stored, shared, and deidentified?
   \answerYes{Yes, see Ethical Considerations}
\end{enumerate}

\end{enumerate}

\appendix
\section*{Appendix}

\begin{figure*}[ht] 
    \centering
    \revfig{\includegraphics[width=\textwidth]{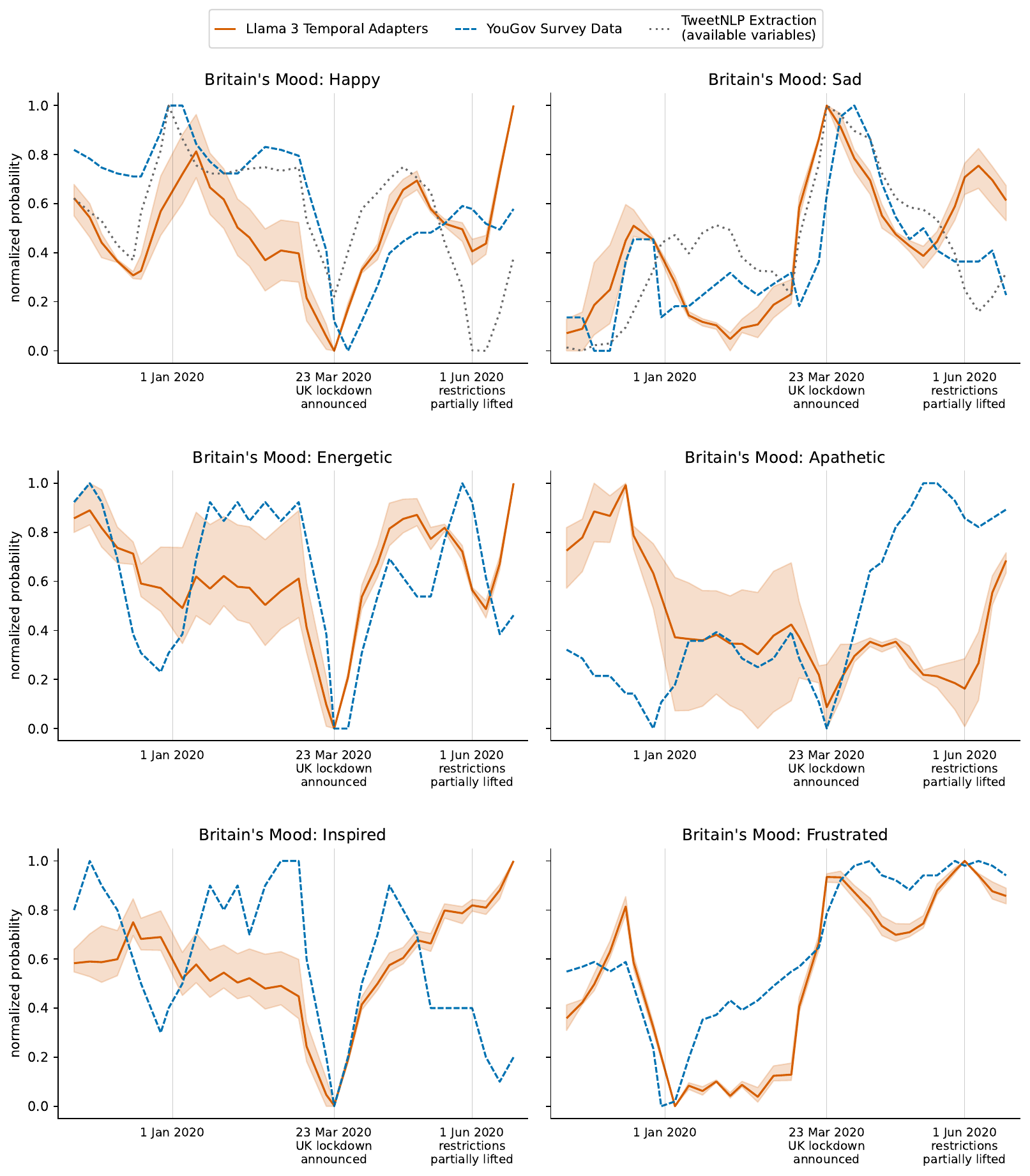}} 
    \caption{\textbf{Time Series of Emotions from Temporal Adapters, Part 1.} We extract answer probabilities by prompting a weekly fine-tuned Llama 3 8B with the same question wording as in the survey~\cite{yougov_britains_2024}, and compare them to the respective weekly survey data. The time series are min-max normalized and a 3 week rolling average is applied. The shaded orange area indicates minimum and maximum LLM answer probabilities across 3 training seeds.}
    \label{fig:fullTimeSeries_part1}
\end{figure*}

\begin{figure*}[ht] 
    \centering
    \revfig{\includegraphics[width=\textwidth]{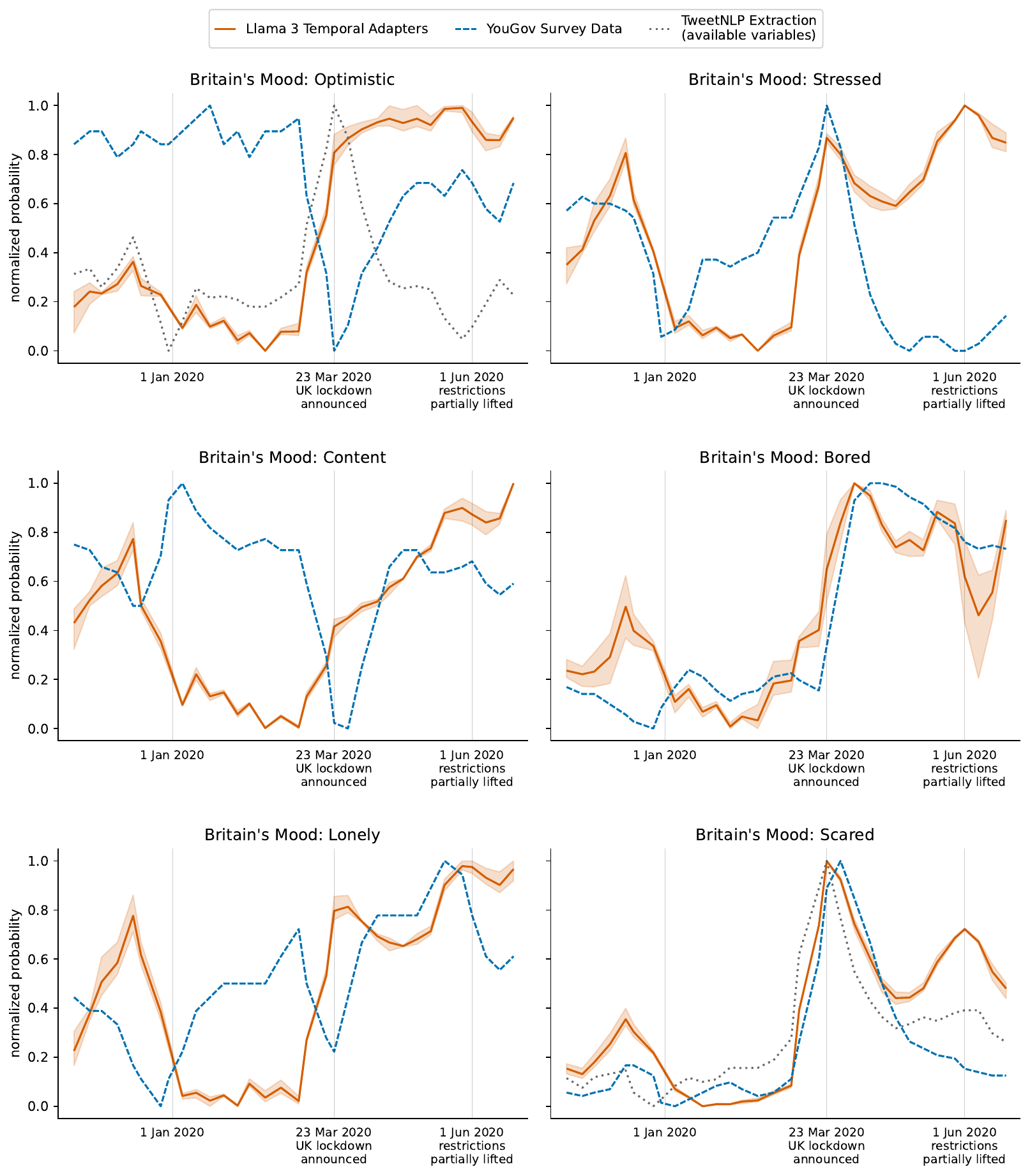}} 
    \caption{\textbf{Time Series of Emotions from Temporal Adapters, Part 2.} We extract answer probabilities by prompting a weekly fine-tuned Llama 3 8B with the same question wording as in the survey~\cite{yougov_britains_2024}, and compare them to the respective weekly survey data. The time series are min-max normalized and a 3 week rolling average is applied. The shaded orange area indicates minimum and maximum LLM answer probabilities across 3 training seeds.}
    \label{fig:fullTimeSeries_part2}
\end{figure*}

\begin{figure*}[ht] 
    \centering
    \revfig{\includegraphics[width=\textwidth]{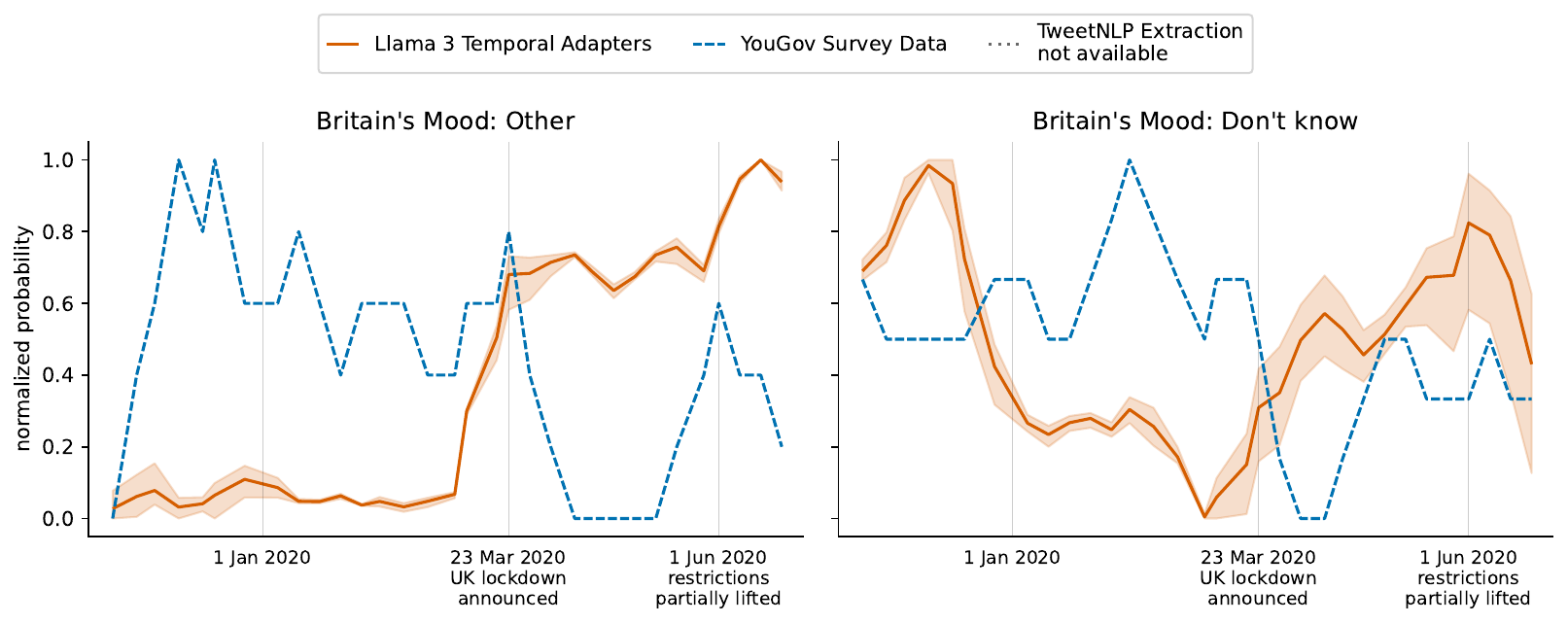}} 
    \caption{\textbf{Time Series of Emotions from Temporal Adapters, Part 3.} We extract answer probabilities by prompting a weekly fine-tuned Llama 3 8B with the same question wording as in the survey~\cite{yougov_britains_2024}, and compare them to the respective weekly survey data. The time series are min-max normalized and a 3 week rolling average is applied. The shaded orange area indicates minimum and maximum LLM answer probabilities across 3 training seeds.}
    \label{fig:fullTimeSeries_part3}
\end{figure*}

\begin{figure*}[ht] 
    \centering
    \revfig{\includegraphics[width=0.95\textwidth]{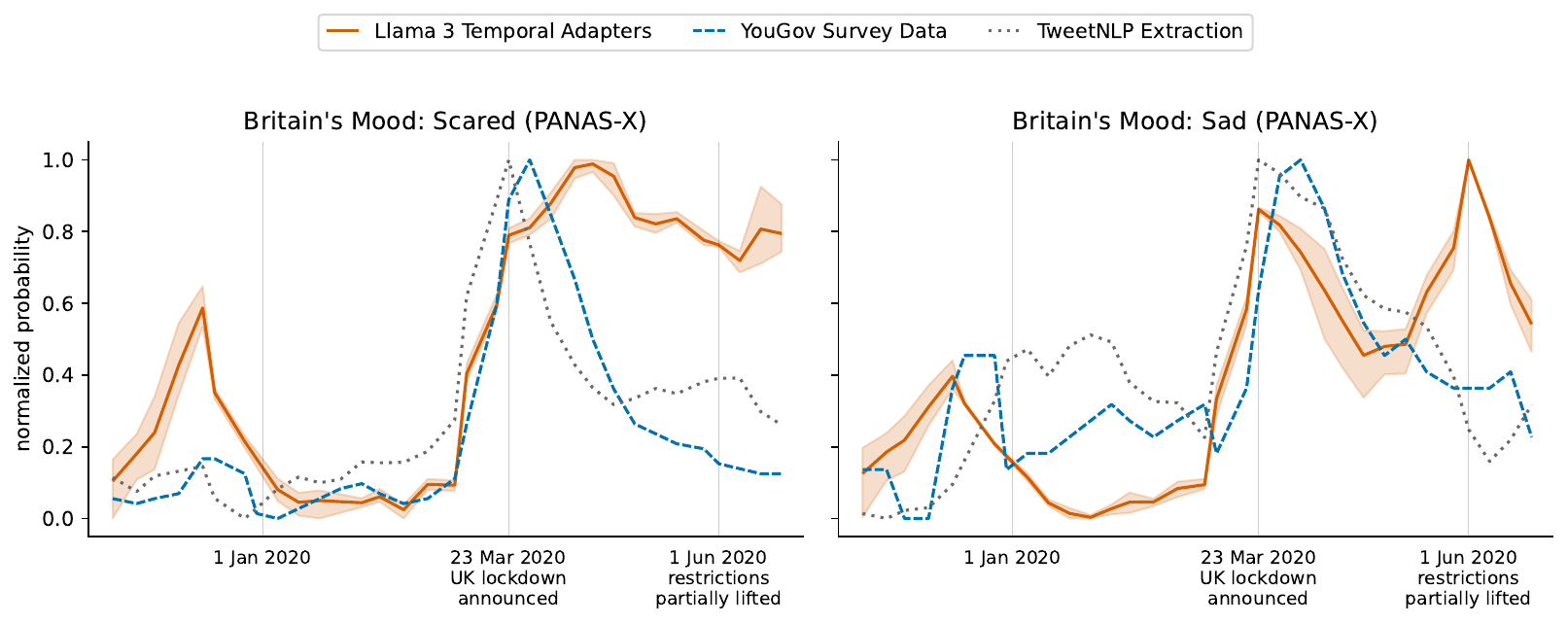}} 
    \caption{\textbf{Time Series of Emotions from Temporal Adapters, Extracted with the PANAS-X Instructions.} We extract answer probabilities by prompting a weekly fine-tuned Llama 3 8B with a question based on the `week' instructions of the PANAS-X inventory~\cite{clark_panas-x_1994}. We combine the responses into a single score as designed by \citet{clark_panas-x_1994}, and compare our results to the respective weekly survey data from~\cite{yougov_britains_2024}. The time series are min-max normalized and a 3 week rolling average is applied. The shaded orange area indicates minimum and maximum LLM answer probabilities across 3 training seeds.}
    \label{fig:fullTimeSeries_PANAS-X}
\end{figure*}

\begin{figure*}[ht] 
    \centering
    \revfig{\includegraphics[width=\textwidth]{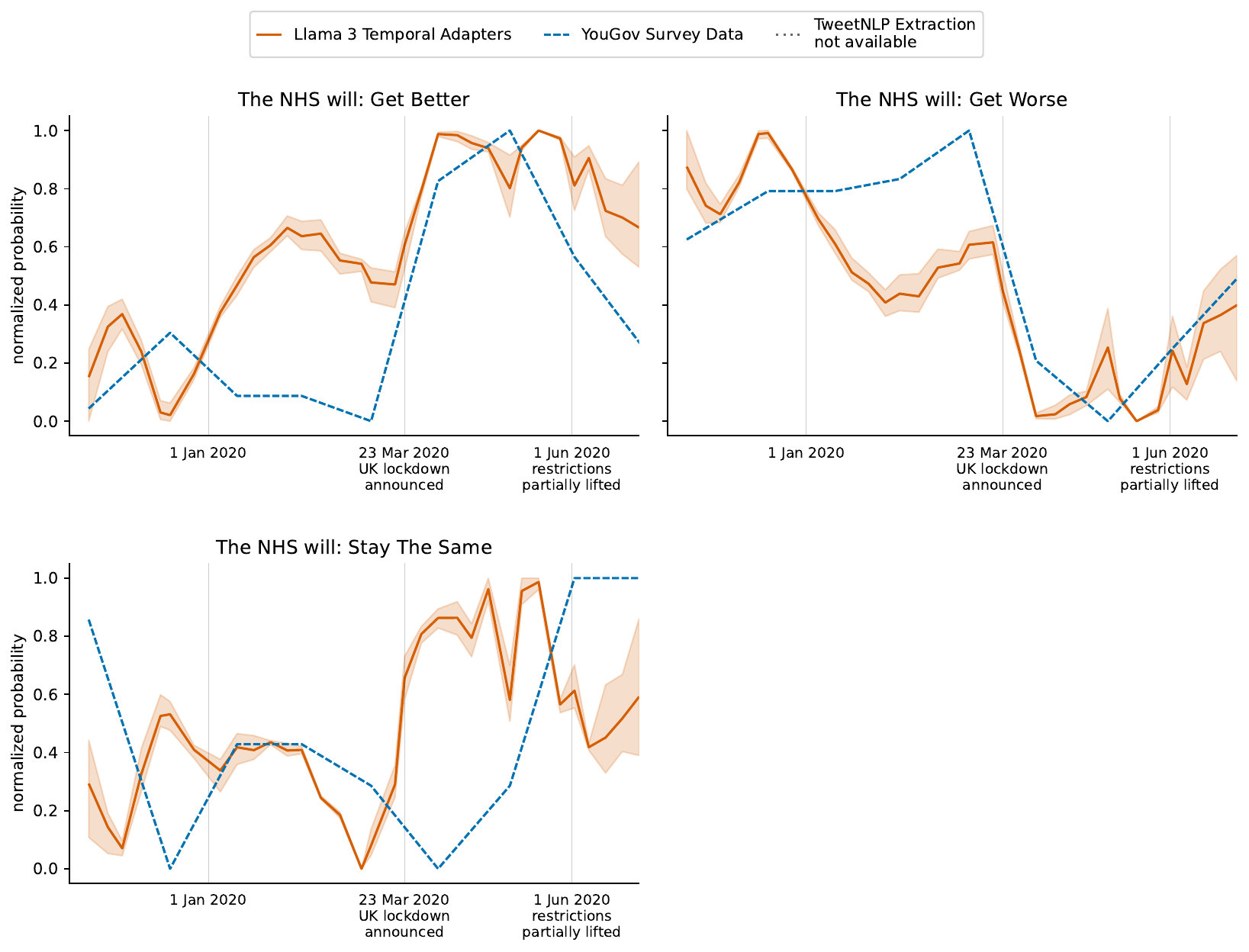}} 
    \caption{\textbf{Estimates of Attitudes towards the \textit{National Health Service} from Temporal Adapters.} We extract answer probabilities by prompting a weekly fine-tuned Llama 3 8B with the same question wording as in the survey~\cite{yougov_will_2024}, and compare them to the respective weekly survey data. The time series are min-max normalized and a 3 week rolling average is applied. The shaded orange area indicates minimum and maximum LLM answer probabilities across 3 training seeds. \citeauthor{yougov_will_2024} survey data for comparison is only available in \textit{monthly} waves. \revision{To the best of our knowledge, no pre-trained classifier is available for labeling tweets with regards to expressed attitudes towards the NHS.}}
    \label{fig:fullTimeSeries_NHSbetter}
\end{figure*}

\begin{figure*}[ht] 
    \centering
    \revfig{\includegraphics[width=\textwidth]{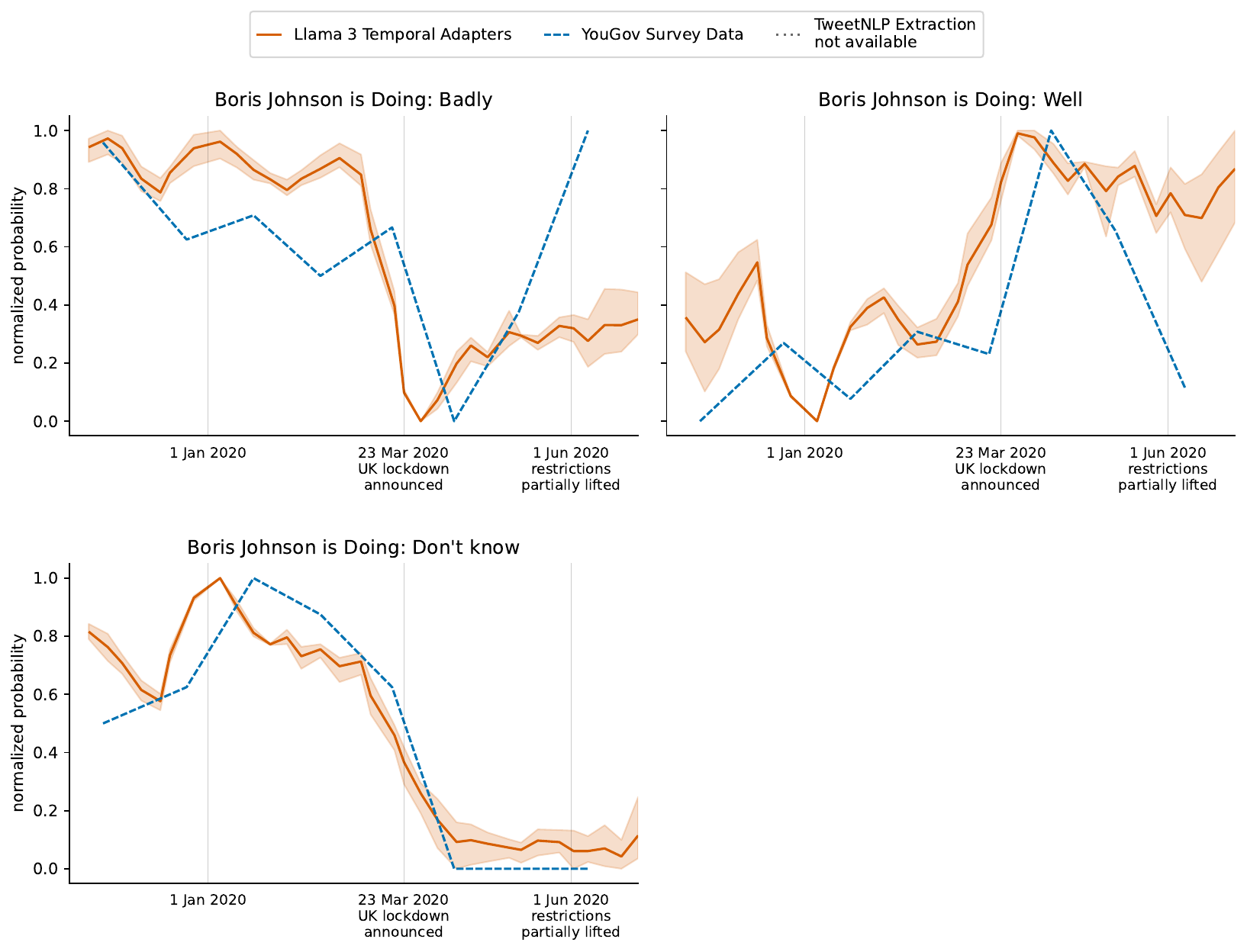}} 
    \caption{\revision{\textbf{Estimates of Attitudes towards \textit{Boris Johnson as Prime Minister} from Temporal Adapters.} We extract answer probabilities by prompting a weekly fine-tuned Llama 3 8B with the same question wording as in the survey~\cite{yougov_how_2024}, and compare them to the respective weekly survey data. The time series are min-max normalized and a 3 week rolling average is applied. The shaded orange area indicates minimum and maximum LLM answer probabilities across 3 training seeds. \citeauthor{yougov_how_2024} survey data for comparison is only available in \textit{monthly} waves. To the best of our knowledge, no pre-trained classifier is available for labeling tweets with regards to this survey question.}}
    \label{fig:fullTimeSeries_BorisPM}
\end{figure*}

\begin{figure*}[ht] 
    \centering
    \revfig{\includegraphics[width=\textwidth]{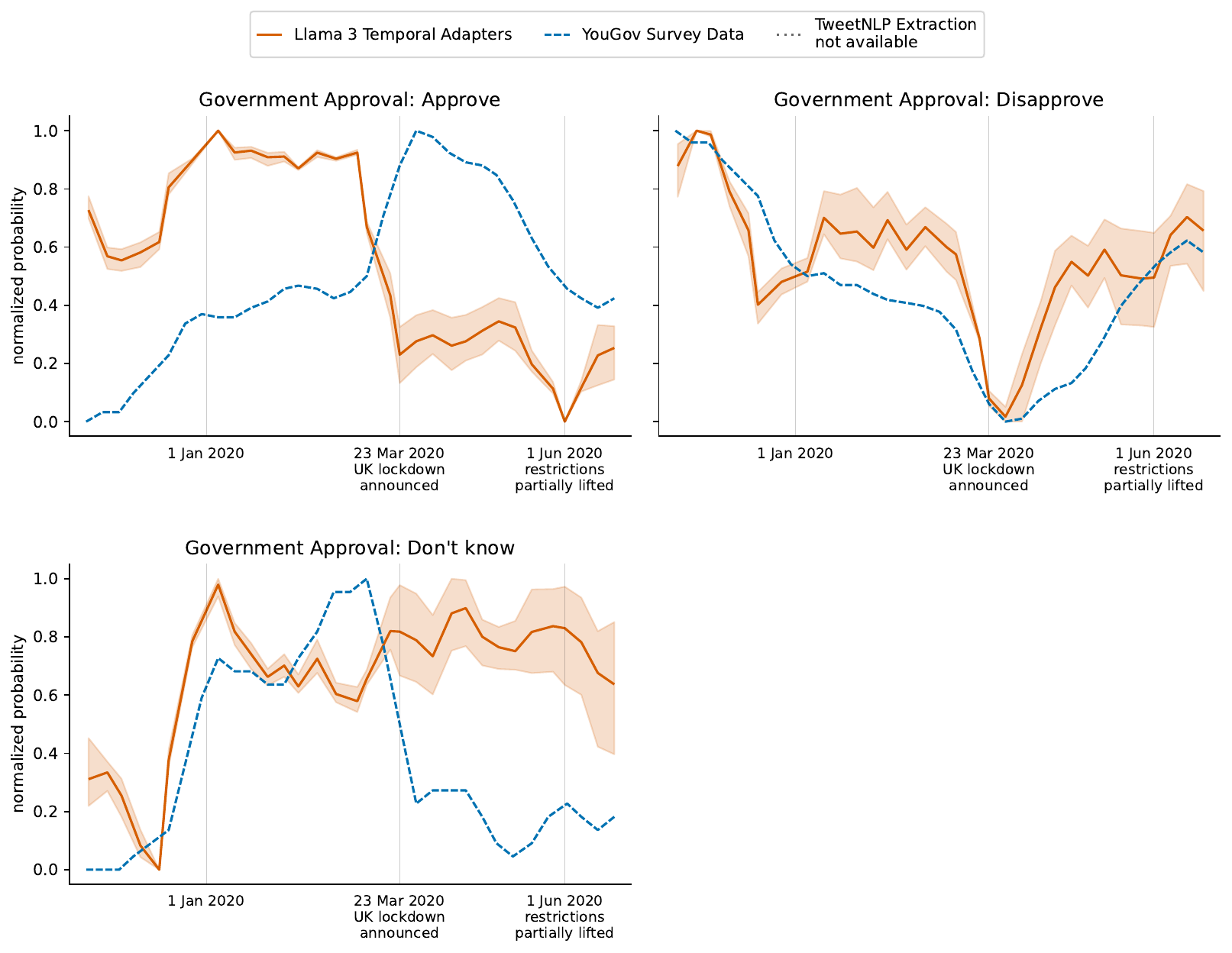}} 
    \caption{\revision{\textbf{Estimates of \textit{Government Approval} from Temporal Adapters.} We extract answer probabilities by prompting a weekly fine-tuned Llama 3 8B with the same question wording as in the survey~\cite{yougov_government_2024}, and compare them to the respective weekly survey data. The time series are min-max normalized and a 3 week rolling average is applied. The shaded orange area indicates minimum and maximum LLM answer probabilities across 3 training seeds. To the best of our knowledge, no pre-trained classifier is available for labeling tweets with regards to this survey question.}}
    \label{fig:fullTimeSeries_govApproval}
\end{figure*}

\begin{figure*}[ht] 
    \centering
    \revfig{\includegraphics[width=0.95\textwidth]{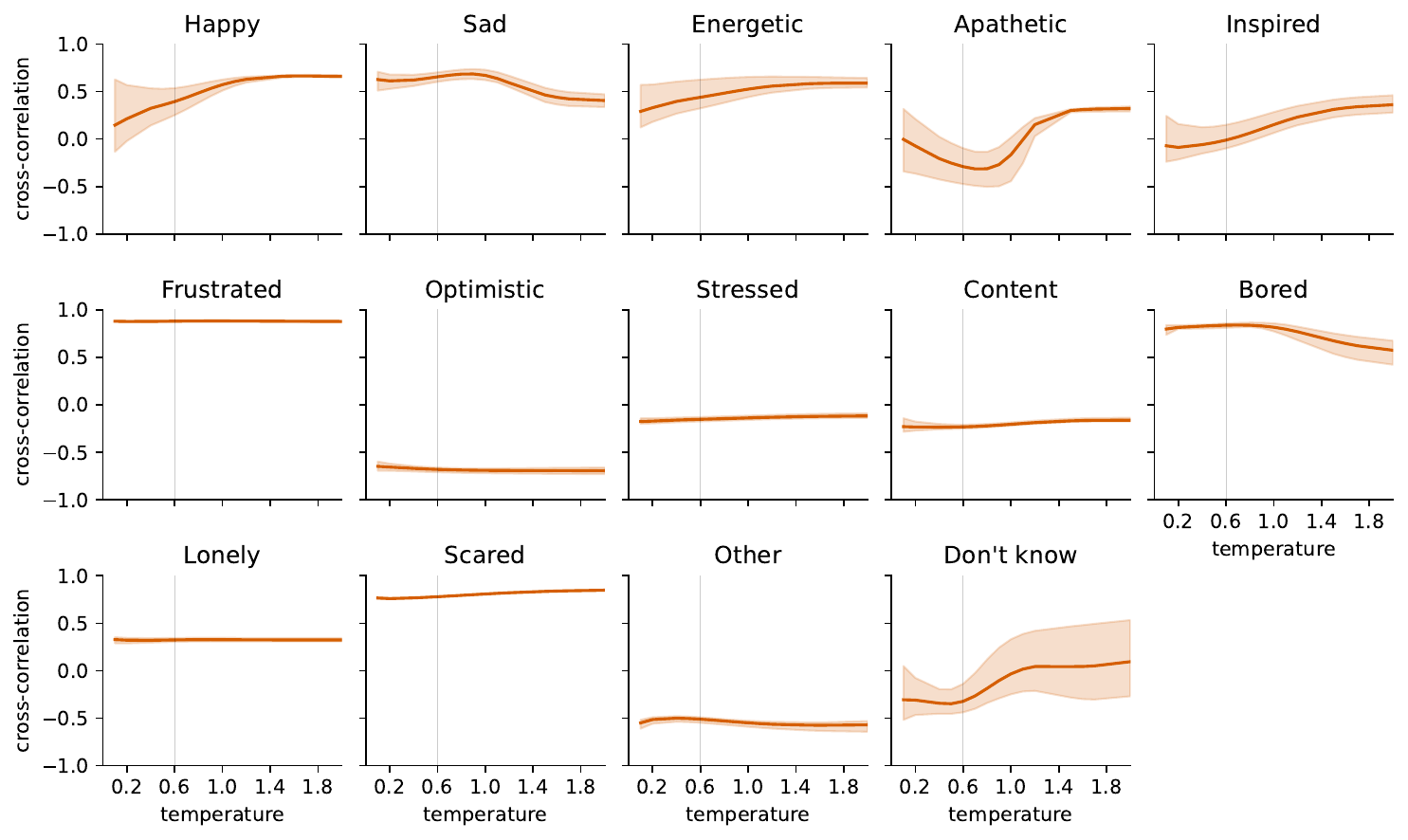}} 
    \caption{\revision{\textbf{Britain`s Mood: Pearson cross-correlations between the YouGov survey data and our estimates, by LLM temperature.} The shaded orange area indicates minimum and maximum cross-correlation across 3 training seeds. We observe that the cross-correlation of many, but not all, answer options is stable across a wide range of LLM temperatures. In this paper, we estimate survey answers based on token probabilities and, unless otherwise specified, use the Llama 3 default temperature of $0.6$ for answer extraction.}}
    \label{fig:corr_by_temp}
\end{figure*}

\begin{figure*}[ht]
    \centering
    \begin{subfigure}{0.265\textwidth}
        \includegraphics[width=\textwidth]{figures/syntheticMix_50steps_v2}
        \caption{50 training steps}
    \end{subfigure}
    \begin{subfigure}{0.265\textwidth}
        \includegraphics[width=\textwidth]{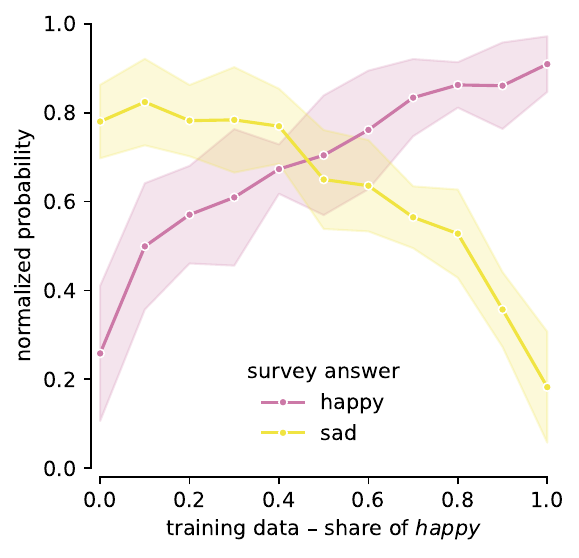}
        \caption{100 training steps}
    \end{subfigure}
    \begin{subfigure}{0.265\textwidth}
        \includegraphics[width=\textwidth]{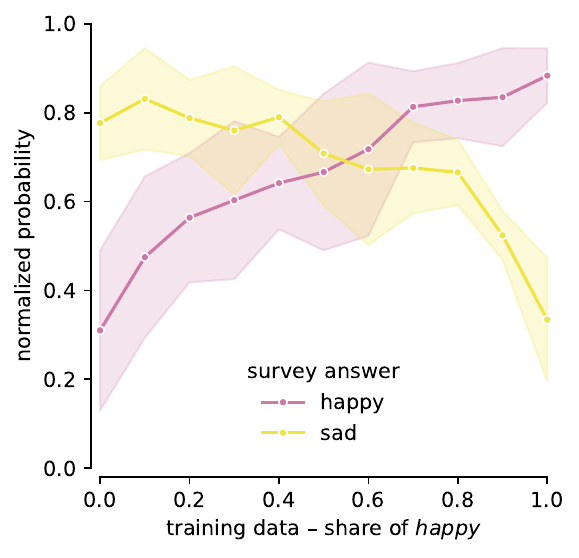}
        \caption{150 training steps}
    \end{subfigure}
    \\
    \begin{subfigure}{0.265\textwidth}
        \includegraphics[width=\textwidth]{figures/syntheticMix_50steps_v2}
        \caption{learning rate $5*10^{-6}$}
    \end{subfigure}
    \begin{subfigure}{0.265\textwidth}
        \includegraphics[width=\textwidth]{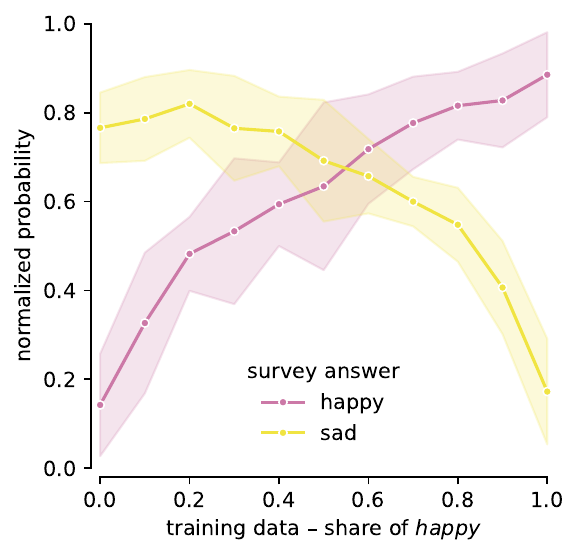}
        \caption{learning rate $10^{-4}$}
    \end{subfigure}
    \\
    \begin{subfigure}{0.265\textwidth}
        \includegraphics[width=\textwidth]{figures/syntheticMix_50steps_v2}
        \caption{temperature $1$}
    \end{subfigure}
    \begin{subfigure}{0.265\textwidth}
        \includegraphics[width=\textwidth]{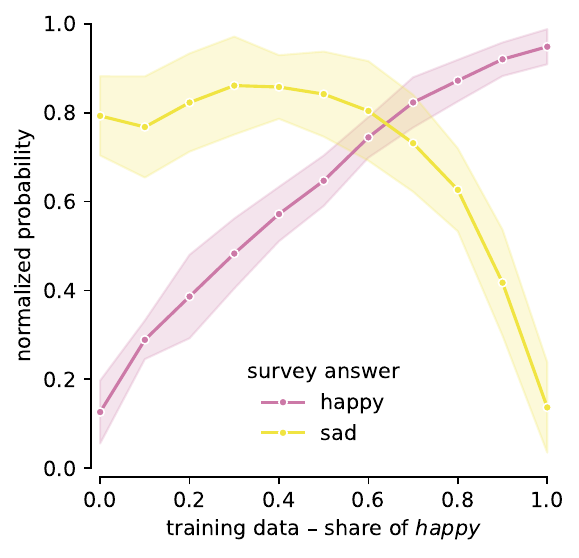}
        \caption{temperature $0.25$}
    \end{subfigure}
    \begin{subfigure}{0.265\textwidth}
        \includegraphics[width=\textwidth]{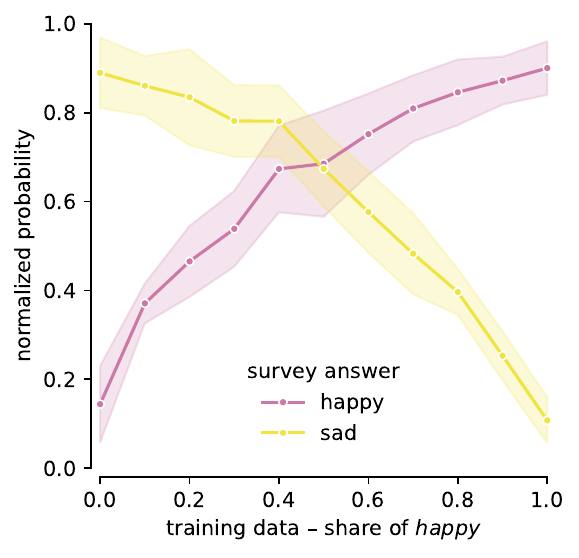}
        \caption{temperature $4$}
    \end{subfigure}
        \\
    \begin{subfigure}{0.265\textwidth}
        \includegraphics[width=\textwidth]{figures/syntheticMix_50steps_v2}
        \caption{answer prefix, answer lowercase}
    \end{subfigure}
    \begin{subfigure}{0.265\textwidth}
        \includegraphics[width=\textwidth]{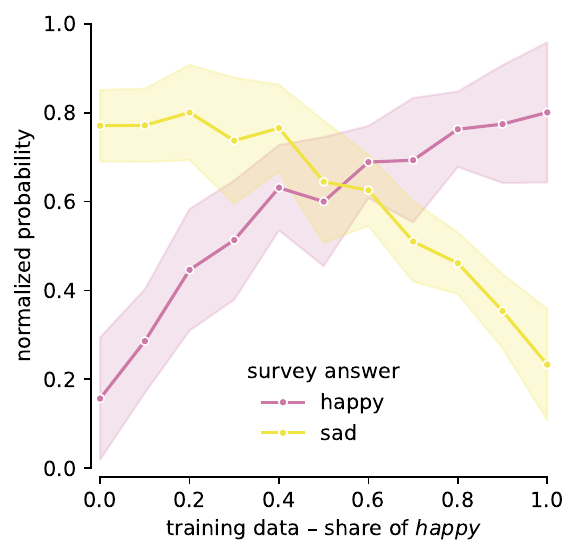}
        \caption{no prefix, answer lowercase}
    \end{subfigure}
    \begin{subfigure}{0.265\textwidth}
        \includegraphics[width=\textwidth]{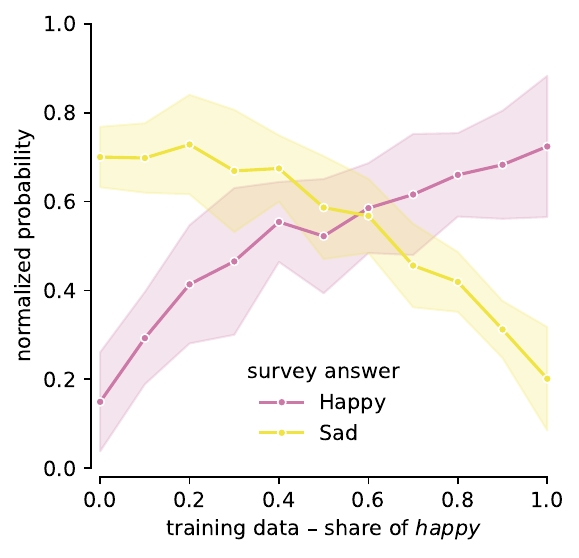}
        \caption{no prefix, answer uppercase}
    \end{subfigure}
    \caption{\textbf{Results from the Partial Hyperparameter Search.} We synthetically mix LLM training data with splits ranging from data that is labeled $100\%$ sad to $100\%$ happy. We then extract answers to \citet{yougov_britains_2024}'s survey question at each split, and show mean and standard deviation over 10 training seeds. Unless otherwise noted, each plot shows results after 50 training steps, with learning rate $5*10^{-6}$, temperature $1$, and using an answer prefix for answer extraction. We aim at a linear relationship between training mix and extracted answers, with low random error across training seeds.}
    \label{fig:hyperparams}
\end{figure*}

\end{document}